\documentclass[11pt]{combine}
\usepackage{amsmath,amsfonts,amssymb,amsthm,epsfig, graphicx, hyperref}
\usepackage[dvipsnames,table,dvipsnames*, svgnames*, hyperref]{xcolor}
\usepackage{natbib,amsmath,amssymb,amsthm,graphicx,setspace,paralist,booktabs,rotating,subcaption,float,color}
\usepackage{multirow}
\usepackage{fancyhdr}
\usepackage{bibunits}
\usepackage[small]{titlesec}

\newtheorem{theorem}{Theorem}
\newtheorem{corollary}{Corollary}
\newtheorem{proposition}{Proposition}

\newtheorem{lemma}{Lemma}
{
	\theoremstyle{definition}
	\newtheorem{definition}{Definition}
	\newtheorem{example}{Example}

}

\newcommand{\beq}{\begin{equation}}
\newcommand{\eeq}{\end{equation}}
\newcommand{\beas}{\begin{eqnarray*}}
	\newcommand{\eeas}{\end{eqnarray*}}
\newcommand{\bea}{\begin{eqnarray}}
\newcommand{\eea}{\end{eqnarray}}
\newcommand{\bei}{\begin{itemize}}
	\newcommand{\eei}{\end{itemize}}
\newcommand{\ben}{\begin{enumerate}}
	\newcommand{\een}{\end{enumerate}}
\newcommand{\bet}{\begin{theorem}}
	\newcommand{\eet}{\end{theorem}}
\newcommand{\bel}{\begin{lemma}}
	\newcommand{\eel}{\end{lemma}}
\newcommand{\bep}{\begin{proposition}}
	\newcommand{\eep}{\end{proposition}}
\newcommand{\bed}{\begin{definition}}
	\newcommand{\eed}{\end{definition}}
\newcommand{\bec}{\begin{corollary}}
	\newcommand{\eec}{\end{corollary}}
\newcommand{\bex}{\begin{example}}
	\newcommand{\eex}{\end{example}}

\def\E{\mathbb{E}}

\def\Sig{\boldsymbol{\Sigma}}

\newcommand{\R}{\mathbb{R}}

\def\V{{\rm Var}}
\def\Sig{\boldsymbol{\Sigma}}

\begin{document}


\title{\scshape Optimal Estimation of Simultaneous Signals Using Absolute Inner Product with Applications to Integrative Genomics}
\author{Rong Ma$^1$, T. Tony Cai$^2$ and Hongzhe Li$^1$ \\
Department of Biostatistics, Epidemiology and Informatics$^1$\\
Department of Statistics$^2$\\
University of Pennsylvania\\
Philadelphia, PA 19104}
\date{}
\maketitle
\thispagestyle{empty}

	\begin{abstract}
 Integrating the summary statistics from  genome-wide association study (\textsc{gwas}) and expression quantitative trait loci (e\textsc{qtl}) data provides a powerful way of identifying  the genes whose expression levels are potentially associated with  complex diseases.  A parameter called $T$-score that quantifies the genetic overlap between a gene and the disease phenotype based on the summary statistics is introduced based on the mean values of two Gaussian sequences. Specifically, given two independent samples $\bold{x}_n\sim N(\theta,  \Sig_1)$ and $\bold{y}_n\sim N(\mu,  \Sig_2)$,  the $T$-score is defined as $\sum_{i=1}^n |\theta_i\mu_i|$, a non-smooth functional, which characterizes the amount of shared signals between two absolute normal mean vectors $|\theta|$ and $|\mu|$.  Using approximation theory,  estimators are constructed and shown to be minimax rate-optimal and adaptive over various parameter spaces. Simulation studies demonstrate the superiority of the proposed estimators over existing methods. The method is applied to an integrative analysis of heart failure  genomics datasets and we identify several genes and biological pathways that are potentially causal to human heart failure. 
	\bigskip
	
\end{abstract}

\section{Introduction}

\subsection{Integrating summary data from GWAS and eQTL studies}

Integrative genomics aims to  integrate various biological data sets for systematic discovery of genetic basis that underlies and modifies human disease \citep{giallourakis2005disease}. To realize its full potential in genomic research, methods of both computational efficiency and theoretical guarantee for such integrative analyses are needed in various applications.  This paper proposes a method that combines datasets from genome-wide association studies (\textsc{gwas}) and expression quantitative trait loci (e\textsc{qtl}) studies in order to identify genetically regulated disease genes and to provide  an integrative view of the underlying biological mechanism of complex diseases such as heart failure.  Results from \textsc{gwas}  have revealed that the majority of disease-associated single nucleotide polymorphisms (\textsc{snp}s) lie in non-coding regions of the genome \citep{hindorff2009potential}.  These  \textsc{snp}s likely regulate the expression of a set of downstream genes that may have effects on diseases \citep{nicolae2010trait}. On the other hand,  e\textsc{qtl} studies measure the association between both cis- and trans- \textsc{snp}s and the expression levels of genes, which characterizes how genetic variants  regulate transcriptions.  A key next step in human genetic research is to explore whether these intermediate cellular level e\textsc{qtl} signals are located in the same loci (``colocalize") as \textsc{gwas} signals and potentially mediate  the genetic effects on disease, and to find disease genes whose e\textsc{qtl}  overlap significantly with the set of loci associated with the disease \citep{he2013sherlock}.

This paper focuses on the integrative analysis of the summary statistics of \textsc{gwas} and e\textsc{qtl} studies performed on possibly different set of subjects.  Due to the privacy and confidentiality concerns of \textsc{gwas}/e\textsc{qtl} participants,   the raw genotype data are often not available, instead most of the published papers provide summary statistics that include single \textsc{snp} analysis results such as the estimated effect size, its $p$-value and the minor allele frequency. 
Based on these summary statistics, we propose a method that identifies potential disease genes by measuring their genetic overlaps to the disease. In particular, we propose a gene-specific measure, $T$-score, that characterizes the total amount of simultaneous \textsc{snp} signals that share the same loci in both \textsc{gwas} and e\textsc{qtl} study of a relevant normal tissues. Such a measure enables us to prioritize genes whose expression levels may underlie and modify human disease \citep{zhao2016sparse}. 

Treating \textsc{snp}-specific \textsc{gwas} and e\textsc{qtl} summary $z$-score statistics (as obtained for linear or logistic regression coefficients) as two independent sequences  of Gaussian random variables, we define the parameter $T$-score as the sum of the product of the absolute values of two normal means over a given set of $n$  \textsc{snp}s. Specifically, for any individual gene $g$, we denote $\bold{x}_n^g$ the vector of $z$-scores from e\textsc{qtl} study, and $\bold{y}_n$ the vector of $z$-scores from \textsc{gwas}. We assume $\bold{x}_n^g \sim N(\theta^g, \Sig_1)$ and $\bold{y}_n \sim N(\mu, \Sig_2)$ for some $\theta^g,\mu\in \R^n$ and covariance matrices $\Sig_1,\Sig_2\in \R^{n\times n}$ with unit diagonals. The $T$-score for gene $g$ is then defined as
\beq
\text{$T$-score}(g) = \sum_{i=1}^n |\theta_i^g\mu_i|,
\eeq
where the summation is over a given set of $n$  \textsc{snp}s. The $T$-score quantifies the amount of simultaneous signals contained in two Gaussian mean vectors, regardless of the directions of the signals. Intuitively, a large $T$-score would possibly result from a large number of contributing components $i$'s whose means $\theta^g_i$ and $\mu_i$ are simultaneously large in absolute values. The supports (nonzero coordinates) of the mean vectors $\theta$ (hereafter we omit its dependence on $g$ for simplicity) and $\mu$ are assumed to have sparse overlaps since it has been observed that, for a relatively large set of  \textsc{snp}s, only a small subset of  \textsc{snp}s are associated with both disease and gene expression \citep{he2013sherlock}. By estimating the $T$-scores for all the genes using summary statistics, we would be able to, after proper normalizations that accounts for study sample sizes, the number of \textsc{snp}s and effect sizes (see Section 2.5), identify and prioritize those genetically regulated candidate disease genes. Besides, the $T$-scores can also be used in the Gene Set Enrichment Analysis to identify the disease associated gene sets and pathways, or to quantify the genetic sharing among different complex traits using the \textsc{gwas} summary statistics \citep{Bulik-Sullivan}.

\subsection{Justification of the absolute inner product}

The $T$-score $\sum_{i=1}^n|\theta_i\mu_i|$ measures the overall signal overlap regardless of the directions of the individual signal components. Although there are other quantities such as $\sum_{i=1}^n\theta_i^2\mu_i^2$ that achieve similar purpose, the $T$-score is closely related to the genetic correlation or genetic relatedness that is widely used  in genetic literature \cite{Bulik-Sullivan}.

Suppose $y$ and $w$ are two traits, and for a given \textsc{snp} with genotype score $x$, the marginal regression functions $y_i=\alpha_x+x_i\beta_x+\epsilon_i$ and $w_i=\eta_x+x_i\gamma_x+\delta_i$ hold for some coefficients $(\alpha_x,\beta_x)$ and $(\eta_x,\gamma_x)$, where $\epsilon_i\sim_{i.i.d.} N(0,\sigma_{x1}^2)$ and $\delta_i\sim_{i.i.d.} N(0,\sigma_{x2}^2)$ for $i=1,2,...,N$ observations. For \textsc{gwas} and e\textsc{qtl} data, one can treat $y$ as a phenotype of interest and $w$ as the expression level of a gene.  In the above models, $x_i\beta_x$ and $x_i\gamma_x$ are the sample-specific marginal genetic effects due to \textsc{snp} $x$, and one can calculate their sample covariance as
\beq
\text{Cov}_x=\frac{1}{N}\sum_{i=1}^N (x_i\beta_x-\bar{x}\beta_x) (x_i\gamma_x-\bar{x}\gamma_x)=\beta_x\gamma_x\cdot \frac{1}{N}\sum_{i=1}^N(x_i-\bar{x})^2,
\eeq
where $\bar{x}=N^{-1}\sum_{i=1}^Nx_i$.
On the other hand, suppose for simplicity that the noise variances $\sigma_{x1}^2$ and $\sigma_{x2}^2$  are known, then the $z$-scores based on the least square estimators $\hat{\beta}_x$ and $\hat{\gamma}_x$ satisfy $$Z_{x1}=\frac{\hat{\beta}_x}{\sigma_{x1}/\sqrt{\sum_{i=1}^N(x_i-\bar{x})^2}}\sim N\left(\frac{{\beta}_x}{\sigma_{x1}/\sqrt{\sum_{i=1}^N(x_i-\bar{x})^2}},1\right)$$ and 
$$Z_{x2}=\frac{\hat{\gamma}_x}{\sigma_{x2}/\sqrt{\sum_{i=1}^N(x_i-\bar{x})^2}}\sim N\left(\frac{{\gamma}_x}{\sigma_{x2}/\sqrt{\sum_{i=1}^N(x_i-\bar{x})^2}},1\right).$$
The product of the mean values of the above $z$-scores satisfies
\beq
\E Z_{x1}\E Z_{x2}=\frac{{\beta_x\gamma_x}}{\sigma_{x1}\sigma_{x2}/{\sum_{i=1}^N(x_i-\bar{x})^2}}=\frac{ \text{Cov}_x}{\sigma_{x1}\sigma_{x2}}.
\eeq
Therefore, in relation to the Gausssian sequence model considered in this paper, the $T$-score is a parameter measuring the sum of absolute normalized sample covariances between the marginal genetic effects across a set of $n$ \textsc{snp}s, i.e., for a set $S$ of \textsc{snp}s, the corresponding $T$-score satisfies
\beq
T\text{-score}=\sum_{x\in S} |\E Z_{x1}\E Z_{x2}| = \sum_{x\in S}{|\text{Cov}_x|}/{(\sigma_{x1}\sigma_{x2})},
\eeq
which measures the overall simultaneous genetic effect of the \textsc{snp}s in $S$.

\subsection{Related works}

Statistically, estimation of $T$-score involves estimating a non-smooth functional -- the absolute value function -- of Gaussian random variables. Unlike the problems of estimating smooth functionals such as the  linear or quadratic functionals \citep{ibragimov1985nonparametric,donoho1990minimax, fan1991estimation,efromovich1994adaptive, cai2005nonquadratic, cai2006optimal} where some natural unbiased estimators are available, much less is known for estimating the non-smooth functionals. Using approximation theory, \cite{cai2011testing} established the minimax risk and constructed a minimax optimal procedure for estimating a non-smooth functional. More recently, this idea has been adapted to statistical information theory that also considered estimation of non-smooth functionals such as the R\'enyi entropy, support size, and $L_1$-norm \citep{jiao2015minimax, jiao2016minimax,wu2016minimax,wu2019chebyshev,acharya2016unified}. In particular, \cite{collier2020estimation} obtained sharp minimax rates for estimating $L_\gamma$-norm for $\gamma\le 1$ under a single sparse Gaussian sequence model, where the optimal rates are achieved by estimators depending on the knowledge of the underlying sparsity. Nonetheless, it remains unknown how to estimate the absolute inner product of two Gaussian mean vectors ($T$-score) with sparse overlap as adaptive as possible.

In the statistical genetics and genomics literature, several  approaches have been proposed for integrating \textsc{gwas} and e\textsc{qtl} data sets. Under the colocalization framework, methods such as \cite{nica2010candidate} and \cite{giambartolomei2014bayesian} were developed to detect colocalised  \textsc{snp}s. However, these methods do not directly identify the potential causal genes. Under the transcriptome-wise association study (TWAS) framework, \cite{zhu2016integration} proposed a summary data-based Mendelian randomization method for causal gene identification, by posing some structural causality assumptions. \cite{Pediscan} developed a gene-based association method called PrediXcan that directly tests the molecular mechanisms through which genetic variation affects phenotype. Nevertheless, there is still a need for a quantitative measure of the genetic sharing between genes and the disease that can be estimated from the \textsc{gwas}/e\textsc{qtl} summary statistics. 

As a related but different quantity, the genetic covariance $\rho$, proposed by \cite{Bulik-Sullivan} as a measure of the genetic sharing between two traits, can be expressed using our notation as  $\rho = \sum_{i=1}^n \theta_i\mu_i$. In addition to the difference due to the absolute value function, in the original definition of genetic covariance $\rho$, the mean vectors $\theta$ and $\mu$ represent the conditional effect sizes (i.e., conditional on all other \textsc{snp}s in the genome), whereas the mean vectors in our $T$-score correspond to the marginal effect sizes, so as to be directly applicable to the standard \textsc{gwas}/e\textsc{qtl} summary statistics (see also \cite{zhao2019cross} for recent developments under high-dimensional linear regression models). In addition, unlike the \textsc{ld}-score regression approach considered in \cite{Bulik-Sullivan}, our proposed method  takes advantage of the fact that the support overlap between $\theta$ and $\mu$ are expected to be very sparse.

\subsection{Main contributions}

In this paper, we propose an estimator of the $T$-score, based on the idea of thresholding and truncating the best polynomial approximation estimator. To the best of our knowledge, this is the first result concerning estimation of such absolute inner product of two Gaussian mean vectors. Under the framework of statistical decision theory, the minimax lower bounds are obtained and we show that our proposed estimators are minimax rate-optimal  over various parameter spaces. In addition, our results indicate that the proposed estimators are locally adaptive to the unknown sparsity level and the signal strength (Section 2). Our simulation study shows the strong empirical performance and robustness of the proposed estimators in various settings, and provides guidelines for using our proposed estimators in practice (Section 3). Analysis of \textsc{gwas} and e\textsc{qtl} data sets of heart failure using the proposed method identifies several  important  genes that are functionally relevant to the etiology of human heart failure (Section 4).

\section{Minimax Optimal Estimation of $T$-score}

\subsection{Minimax lower bounds}

We start with establishing the minimax lower bounds for estimating $T$-score over various parameter spaces. Throughout, we denote $T(\theta,\mu)=\sum_{i=1}^n|\theta_i\mu_i|$. For a vector ${a} = (a_1,...,a_n)^\top \in \mathbb{R}^{n}$, we define  $\| {a}\|_{\infty} = \max_{1\le j\le n}  |a_{i}|$. For sequences $\{a_n\}$ and $\{b_n\}$, we write $a_n\lesssim b_n$ or $b_n \gtrsim a_n$ if there exists an absolute constant $C$ such that $a_n \le Cb_n$ for all $n$, and write $a_n\asymp b_n$ if $a_n \lesssim b_n$ and $a_n\gtrsim b_n$. 

As of both practical and theoretical interest, we focus on the class of mean vector pairs $(\theta,\mu)$  with only a small fraction of support overlaps. 
Specifically,  for any $s< n$, we define the parameter space for $(\theta,\mu)$ as $D(s) =\{(\theta,\mu)\in \R^n\times \R^n : |\text{supp}(\theta) \cap \text{supp}(\mu)| \le s\}.$
Intuitively, in addition to the sparsity $s$, the difficulty of estimating $T(\theta,\mu)$ should also rely on the magnitudes of the mean vectors $\theta$ and $\mu$, and the covariance matrices $\Sig_1$ and $\Sig_2$. Towards this end, we define the parameter space for $(\theta,\mu,\Sig_1,\Sig_2)$ as  $D^\infty(s,L_n) = \big\{(\theta,\mu,\Sig_1,\Sig_2):  (\theta,\mu)\in D(s),\max(\|\theta\|_\infty , \|\mu\|_\infty )\le L_n,  \Sig_1=\Sig_2={\bf I}_n \big\}$,
where both $s$ and $L_n$ can growth with $n$.
In particular, to construct estimators that are as adaptive as possible, and to avoid unnecessary complexities of extra logarithmic terms, we calibrate the sparsity $s \asymp n^{\beta}$ for some $0<\beta<1$. Throughout, we consider the normalized loss function as the squared distance scaled by $n^{-2}$ and define the estimation risk for some estimator $\hat{T}$ as $\mathcal{R}(\hat{T})=\frac{1}{n^2}\E (\hat{T}-T(\theta,\mu))^2$.
To simplify our statement, we define the rate function $\psi(s,n) = \min\left\{ \log\left(1+\frac{n}{s^2}\right),L^2_n\right\}+\frac{\min\{\log s, L_n^2\}}{\log^2 s}.$
The following theorem establishes the minimax lower bound over  $D^{\infty}(s,L_n)$.

\bet \label{sparse.lower3}
Let $\bold{x_n} \sim N(\theta,\Sig_1)$ and $\bold{y_n} \sim N(\mu,\Sig_2)$ be multivariate Gaussian random vectors where $(\theta,\mu,\Sig_1,\Sig_2) \in D^\infty(s,L_n)$. Then 
\begin{equation} \label{sparse.lower.equation4}
\inf_{\hat{T}} \sup_{\substack{(\theta,\mu,\Sig_1,\Sig_2) \in D^\infty(s,L_n)}}	\mathcal{R}(\hat{T}) \gtrsim \frac{L_n^2s^2\psi(s,n)}{n^2}
\end{equation}
where $\hat{T}$ is any estimator based on $(\bold{x_n} ,\bold{y_n} )$.
\eet
From the above theorem and the definition of the rate function $\psi(s,n)$, when $\beta\in (0,1/2)$, (\ref{sparse.lower.equation4}) becomes
\beq \label{lb.1}
\inf_{\hat{T}} \sup_{\substack{(\theta,\mu,\Sig_1,\Sig_2) \in D^{\infty}(s,L_n)}}	\mathcal{R}(\hat{T})\gtrsim\frac{L_n^2s^2}{n^2}\min\{\log n,L_n^2\},
\eeq
when $\beta\in (1/2,1)$, we have
\beq
\inf_{\hat{T}} \sup_{\substack{(\theta,\mu,\Sig_1,\Sig_2) \in D^{\infty}(s,L_n)}} 	\mathcal{R}(\hat{T}) \gtrsim  \frac{L_n^2s^2}{n^2\log^2 n}\min\{\log n,L_n^2\},
\eeq
and when $\beta=1/2$, we have $\inf_{\hat{T}} \sup_{\substack{(\theta,\mu,\Sig_1,\Sig_2) \in D^{\infty}(s,L_n)}} 	\mathcal{R}(\hat{T}) \gtrsim  \frac{L_n^2s^2}{n^2}.$

\subsection{Optimal estimators of $T$-score via polynomial approximation}

In general, the proposed estimators are based on the idea of optimal estimation of the absolute value of normal means studied  by \cite{cai2011testing}.  Therein, the best polynomial approximation of the absolute value function was applied to obtain the optimal estimator and the minimax lower bound. Specifically, it was shown that, if we define the $2K$-degree polynomial 
$
G_K(x) = {2 \over \pi}T_0(x) +{4 \over \pi}\sum_{k=1}^{K} (-1)^{k+1}
{T_{2k}(x) \over {4k^2-1}} \equiv \sum_{k=0}^K g_{2k} x^{2k},
$
where $T_{k}(x) = \sum_{j=0}^{[{k/2}]} (-1)^j {k\over k - j} {k-j\choose j} 2^{k-2j-1}
x^{k-2j}$ are Chebyshev polynomials, 
then for any $X \sim N(\theta,1)$, if $H_k$ are Hermite polynomials with respect to the standard normal density $\phi$ such that ${d^k \over dy^k}\phi(y) = (-1)^k H_k(y) \phi(y)$,
the estimator based on $\tilde{S}_K(X) \equiv \sum_{k=0}^K g_{2k}M_n^{-2k+1}H_{2k}(X)$
for some properly chosen $K$ and $M_n$ has some optimality properties for estimating $|\theta|$. This important result motivates our construction of the optimal estimators of $T$-score. 

We begin by considering the setting where  $\bold{x}_n=(x_1,...,x_n)^\top \sim N(\theta,{\bf I}_n)$ and $\bold{y}_n=(y_1,...,y_n)^\top \sim N(\mu,{\bf I}_n)$. To estimate $T(\theta,\mu)$, we first split each sample into two copies, one is used for testing, and the other is used for estimation. Specifically, for $x_i \sim N(\theta_i,1)$, we generate $x_{i1}$ and $x_{i2}$ from $x_{i}$ by letting $z_i \sim N(0,1)$ and setting $x'_{i1} = x_i+z_i$ and $x'_{i2} = x_i-z_i$. Let $x_{il} =  x'_{il}/\sqrt{2}$ for $l=1,2$, then $x_{il} \sim N(\theta'_i,1)$ for $l=1,2$ and $i=1,...,n$ with $\theta_i' = \theta_i/\sqrt{2}$. Similarly, we construct $y_{il}\sim N(\mu'_i,1)$ for $l=1,2$ and $i=1,...,n$ with $\mu'_i = \mu_i/\sqrt{2}$. Since $T(\theta,\mu)=2T(\theta',\mu')$, estimating $T(\theta,\mu)$ with $\{ x_i,y_i\}_{i=1}^n$ is equivalent to estimating $T(\theta', \mu')$ with $\{x_{il},y_{il} \}_{i=1}^n, l=1,2$. 

In light of the estimator $\tilde{S}_K(X) $, we consider a slightly adjusted statistic $S_K(X) = \sum_{k=1}^K g_{2k}M_n^{-2k+1}H_{2k}(X)$ and define its truncated version ${\delta}_K(X) = \min\{{S}_K(X) ,n^2 \}$,
with $M_n = 8\sqrt{\log n}$ and $K\ge 1$ to be specified later. The above truncation is important in reducing the variance of $\delta_K(X)$.
By the sample splitting procedure, we construct an estimator of $|\theta'_i|$  as
\[
\hat{V}_{i,K}(x_i)= {\delta}_K(x_{i1})I(|x_{i2}|\le 2\sqrt{2\log n}) + |x_{i1}| I(|x_{i2}|> 2\sqrt{2\log n}),
\]
and a similar estimator of $|\mu'_i|$ as $\hat{V}_{i,K}(y_i)$. To further exploit the sparse structure, we also consider their thresholded version 
\[
\hat{V}^S_{i,K}(x_i)= {\delta}_K(x_{i1})I(\sqrt{2\log n}<|x_{i2}|\le 2\sqrt{2\log n}) + |x_{i1}| I(|x_{i2}|> 2\sqrt{2\log n})
\]
as an estimator of $|\theta'_i|$ and similarly $\hat{V}^S_{i,K}(y_i)$ for $|\mu'|$. Intuitively, both $\hat{V}_{i,K}(x_i)$ and $\hat{V}^S_{i,K}(x_i)$ are hybrid estimators: $\hat{V}_{i,K}(x_i)$ is a composition of an estimator based on polynomial approximation designed for small to moderate observations (less than $2\sqrt{2\log n}$ in absolute value) and the simple absolute value estimator applied to large observations (larger than $2\sqrt{2\log n}$ in absolute value), whereas $\hat{V}^S_{i,K}(x_i)$ has an additional thresholding procedure for small observations (less than $\sqrt{2\log n}$ in absolute value). Consequently, we propose two estimators of $T(\theta,\mu)$, namely,
\begin{equation} \label{sparse.est}
\widehat{T}_K = {2}\sum_{i=1}^n \hat{V}_{i,K}(x_i)\hat{V}_{i,K}(y_i),
\end{equation}
as the hybrid non-thresholding estimator and
\beq
\widehat{T}^S_K= {2}\sum_{i=1}^n \hat{V}^S_{i,K}(x_i)\hat{V}^S_{i,K}(y_i),
\eeq
as the hybrid thresholding estimator. Both estimators rely on $K$, which is the tuning parameter to be specified later.

\subsection{Theoretical properties and minimax optimality}

The following theorem provides the risk upper bounds of $\widehat{T}_K$ and $\widehat{T}^S_K$ over $D^{\infty}(s,L_n)$.

\bet \label{sparse.upper3}
Let $\bold{x_n} \sim N(\theta,\Sig_1)$ and $\bold{y_n} \sim N(\mu,\Sig_2)$ be multivariate Gaussian random vectors with $(\theta,\mu,\Sig_1,\Sig_2)\in D^{\infty}(s,L_n)$ and $s\asymp n^\beta$. Then 
\begin{enumerate}
	\item for any $\beta\in(0,1)$ and $K$ being any finite positive integer, we have
	\begin{equation} \label{sparse.upper.equation5}
	\sup_{(\theta,\mu,\Sig_1,\Sig_2)\in D^{\infty}(s,L_n)}	\mathcal{R}(\widehat{T}^S_K ) \lesssim \frac{(L^2_n+\log n)s^2\log n}{n^2};
	\end{equation}
	if in addition $L_n\le (\sqrt{2}-1)\sqrt{\log n}$, then
	\beq \label{ub.2}
	\sup_{(\theta,\mu,\Sig_1,\Sig_2)\in D^{\infty}(s,L_n)}	\mathcal{R}(\widehat{T}^S_K )\lesssim \frac{s^2L_n^4}{n^2}+\frac{\log^2 n}{n^{5/2}}+\frac{L_n^2\log n}{n^2};
	\eeq
	\item for any $\beta\in (1/2,1)$ and $K= r\log n$ for some $0<r<\frac{2\beta-1}{12}$, we have
	\begin{equation} \label{sparse.upper.equation6}
	\sup_{(\theta,\mu,\Sig_1,\Sig_2)\in D^{\infty}(s,L_n)}	\mathcal{R}(\widehat{T}_K ) \lesssim \frac{(L^2_n+1/\log n)s^2}{n^2\log n}.
	\end{equation}
\end{enumerate}
\eet

\begin{figure}[h!]
	\centering
	\includegraphics[angle=0,width=10cm]{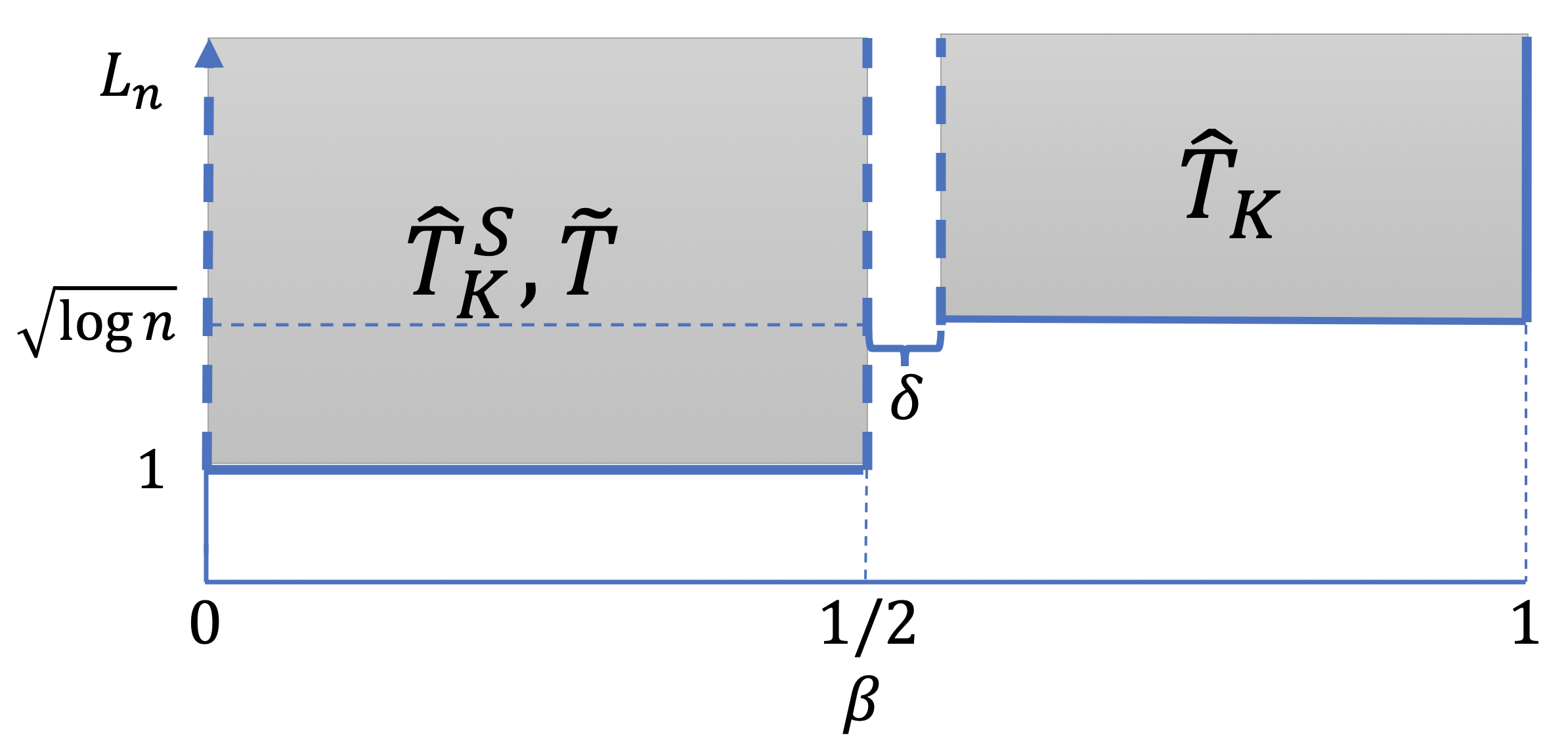}
	\vspace{-0.3cm}
	\caption{A graphical illustration of the regions where the proposed estimators are minimax optimal and adaptive. Among them, $\widehat{T}^S_K$ has $K$ being any finite positive integer and $\widehat{T}_K$ has $K =r\log n$ for some $0<r<\delta/6$.}
	\label{rate}
\end{figure}

Over the sparse region $\beta\in(0,1/2)$, the risk upper bounds (\ref{sparse.upper.equation5}) and (\ref{ub.2}) along with the minimax lower bound (\ref{lb.1}) implies that $\widehat{T}^S_K$ with $K$ being any finite positive integer is minimax rate-optimal over $D^{\infty}(s,L_n)$ when $L_n\gtrsim 1$, where the minimax rate is
\beq \label{r1}
\inf_{\hat{T}} \sup_{\substack{(\theta,\mu,\Sig_1,\Sig_2) \in D^{\infty}(s,L_n)}}	\mathcal{R}(\hat{T})\asymp \frac{L_n^2s^2}{n^2}\min\{\log n, L_n^2\}.
\eeq
When $L_n\lesssim 1$, the problem is less interesting since in this case, the trivial estimator $0$ attains the minimax rate $L_n^4s^2/n^2$.
Over the dense region $\beta\in(1/2,1)$, the non-thresholding estimator $\widehat{T}_K$ with $K=r\log n$ for some small $r$ is minimax rate-optimal over $D^{\infty}(s,L_n)$ for $L_n\gtrsim \sqrt{\log n}$, where the minimax rate is
\beq \label{r2}
\inf_{\hat{T}} \sup_{\substack{(\theta,\mu,\Sig_1,\Sig_2) \in D^{\infty}(s,L_n)}} 	\mathcal{R}(\hat{T}) \asymp  \frac{L_n^2s^2}{n^2\log n}.
\eeq
In both cases, the tuning parameter $K$ plays an important role in controlling the bias-variance trade-off. An important consequence of our results concerns the {local adaptivity} of $\widehat{T}_K$ and $\widehat{T}^S_K$ with respect to $s$ and $L_n$. Specifically, for any $\delta>0$, the estimator $\widehat{T}_K$ with $K =r\log n$ for some $0<r<\delta/6$ is simultaneously rate-optimal for any $L_n\gtrsim \sqrt{\log n}$ and any $\beta\in(1/2+\delta,1)$, whereas the estimator $\widehat{T}_K$ with $K$ being any finite positive integer is simultaneously rate-optimal for any $L_n\gtrsim 1$ and $\beta\in(0,1/2)$.  See Figure \ref{rate} for an illustration.

Unfortunately, even with appropriate choices of $K$, neither $\widehat{T}^S_K$ nor $\widehat{T}_K$ is simultaneously optimal across all $\beta\in(0,1)$. However, since the difference in the optimal rates of convergence between (\ref{r1}) and (\ref{r2}) is only of a factor of $\log n$, we can see that in practice, even when $\beta\in(1/2,1)$, the thresholding estimator $\widehat{T}^S_K$ would perform just as well as the non-thresholding estimator $\widehat{T}_K$. See Section 3 for detailed numerical studies. 

\subsection{Sparse estimation via simple thresholding}

According to our previous analysis, if the parameter space is very sparse, i.e., $\beta\in(0,1/2)$, the proposed estimator $\widehat{T}^S_K$ is minimax optimal if we choose $K$ as any constant positive integer. In other words, any constant degree polynomial approximation suffices for the optimal estimation of $T(\theta,\mu)$, including the constant function.  It means that in this case the polynomial approximation is essentially redundant for our purpose. 

In light of the above observation, we consider the simple thresholding estimator $\widetilde{T}={2}\sum_{i=1}^n \hat{U}_i(x_i)\hat{U}_i(y_i)$, where $\hat{U}_i(x_i)= |x_{i1}| I(|x_{i2}|> 2\sqrt{2\log n})$.
Our next  theorem obtains the risk upper bound of $\widetilde{T}$  over $D^{\infty}(s,L_n)$, which along with (\ref{lb.1}) from Theorem \ref{sparse.lower3} shows that $\widetilde{T}$ is also minimax optimal and adaptive over any sparsity level $\beta\in(0,1/2)$ and $L_n\gtrsim 1$.

\bet \label{sparse.upper.3}
Let $\bold{x_n} \sim N(\theta,\Sig_1)$ and $\bold{y_n} \sim N(\mu,\Sig_2)$ be multivariate Gaussian random vectors with $(\theta,\mu,\Sig_1,\Sig_2) \in D^{\infty}(s,L_n)$. Then
\beq
\sup_{\substack{(\theta,\mu,\Sig_1,\Sig_2) \in D^\infty(s,L_n)}}\mathcal{R}(\widetilde{T} ) \lesssim \frac{(L_n^2+\log n)s^2\log n}{n^2}.
\eeq
If in addition $L_n\le \sqrt{2\log n}$, then
\beq
\sup_{\substack{(\theta,\mu,\Sig_1,\Sig_2) \in D^\infty(s,L_n)}}\mathcal{R}(\widetilde{T} ) \lesssim \frac{s^2L_n^4}{n^2}+\frac{\log^2 n}{n^3}+\frac{L_n^2\log n}{n^2}.
\eeq
\eet

Since our simple thresholding estimator $\widetilde{T}$ completely drops the polynomial components in $\widehat{T}^S_K$, its variance is significantly reduced. As a consequence, we find that as long as $\max(\|\theta\|_\infty , \|\mu\|_\infty )\le \sqrt{n}$, the condition $\Sig_1=\Sig_2={\bf I}_n$ can be removed without changing the rate of convergence. Towards this end, we define the enlarged parameter space
\[
D_0^\infty(s,L_n) = \left\{(\theta,\mu,\Sig_1,\Sig_2): \begin{aligned} & (\theta,\mu)\in D(s),\max(\|\theta\|_\infty , \|\mu\|_\infty )\le L_n, \\
&  \Sig_1,\Sig_2\succeq 0, \text{$\Sig_1$ and $\Sig_2$ have unit diagonals}.
\end{aligned} \right\}.
\]
In particular, as $\Sig_1$ and $\Sig_2$ have unit diagonals, the sample splitting procedure (Section 2.1) still applies, which only leads to a $1/2$-scaling of the off-diagonal entries of the covariance matrices.
\bet \label{sparse.upper.4}
Let $\bold{x_n} \sim N(\theta,\Sig_1)$ and $\bold{y_n} \sim N(\mu,\Sig_2)$ where $(\theta,\mu,\Sig_1,\Sig_2)\in D_0^{\infty}(s, L_n)$ and $L_n\lesssim \sqrt{n}$. Then we have
\beq
\sup_{\substack{(\theta,\mu,\Sig_1,\Sig_2) \in D_0^\infty(s,L_n)}}\mathcal{R}(\widetilde{T} )\lesssim \frac{(L_n^2+\log n)s^2\log n}{n^2}.
\eeq
\eet

By definition, we have $D^{\infty}(s,L_n)\subset D_0^{\infty}(s,L_n)$. It then follows from Theorems \ref{sparse.lower3} and \ref{sparse.upper.4} that for any $\beta\in(0,1/2)$ and $L_n\lesssim \sqrt{n}$
\beq
\inf_{\hat{T}} \sup_{\substack{(\theta,\mu,\Sig_1,\Sig_2) \in D_0^\infty(s,L_n)}}\mathcal{R}(\hat{T}) \asymp \frac{s^2L_n^2}{n^2}\cdot \min\{\log n,L_n^2\},
\eeq
where the minimax optimal rate can be attained by $\widetilde{T}$ when $L_n\ge \sqrt{\log n}$ and by the trivial estimator $0$ when $L_n< \sqrt{\log n}$. This establishes the minimax optimality and adaptivity of $\widetilde{T}$ over $D_0^{\infty}(n^\beta,L_n)$ for any $\beta\in(0,1/2)$ and $L_n\gtrsim \sqrt{\log n}$.
The result confirms an important advantage of $\widetilde{T}$ over $\widehat{T}^S_K$, namely, its guaranteed theoretical performance over arbitrary correlation structures, which complies with the fact that in many applications the observations are not necessarily independent. For more detailed discussions on estimation with non-identity covariances or unknown covariances, see Section A.2 of our Supplementary Material.

\subsection{Normalization, linkage disequilibrium and the use of $T$-score}

Dealing with linkage disequilibrium  (\textsc{ld}) among the  \textsc{snp}s \citep{reich2001linkage,daly2001high,pritchard2001linkage} is essential in any genetic studies. In this paper, we follow the idea of \cite{Bulik-Sullivan} and propose  to use the normalized $T$-score
\[
\text{Normalized $T$-score}(g) = \frac{\sum_{i=1}^n|\theta_i^g\mu_i|}{\|\theta^g\|_2\|\mu\|_2}
\]
as a measure of genetic overlap between gene $g$ and the outcome disease. In particular, the estimation of the $\ell_2$ norms $\|\theta^g\|_2$ and $\|\mu\|_2$, or in our context, the \textsc{snp}-heritability of the traits \citep{yang2010common}, can be easily accomplished using summary statistics. As a result, every normalized $T$-score lies between $0$ and $1$, which is scale-invariant (e.g., invariance to study sample sizes and \textsc{snp} effect sizes) and comparable across many different genes or studies. In addition, as argued by \cite{Bulik-Sullivan}, the normalized $T$-score is less sensitive  to  the choice of the  $n$-\textsc{snp} sets.

Moreover, in Theorem 4, we show that the simple thresholding estimator $\widetilde{T}$ does not require the independence of the $z$-scores, which theoretically guarantees its applicability in the presence of arbitrary \textsc{ld} structure among the  \textsc{snp}s. However, our theoretical results concerning $\widehat{T}_K$ and $\widehat{T}^S_K$ rely on such an  independence assumption.  In our simulation studies, we found that the empirical performance (including optimality) of $\widehat{T}_K$ and $\widehat{T}^S_K$ is not likely affected by the dependence due to the  \textsc{ld} structure. As a result, our proposed estimation method, although partially analysed under the independence assumption, can be directly applied to the summary statistics without specifying the underlying \textsc{ld} or covariance structure.

The $T$-score can be used for identifying disease genes and pathways using the \textsc{gwas} and e\textsc{qtl}  data.  For each gene, we estimate the $T$-score by our proposed estimators using the vectors of $z$-scores from \textsc{gwas} and e\textsc{qtl} studies. After obtaining the estimated $T$-scores for all the genes and the corresponding \textsc{snp}-heritability, we rank the genes by the order of their normalized $T$-scores. As a result, genes with the highest ranks are considered important in gaining insights into  the biological mechanisms of the diseases.  For gene set or pathway analysis, we obtain the normalized $T$-scores $T_j$, $1\le j\le J$ for given a gene set $S$ and then calculate the Kolmogorov-Smirnov test  statistic defined as $\sup_t | \frac{1}{k}\sum_{j\in S}I(T_j \le t) -\frac{1}{k'}\sum_{j\in S^c}I(T_j \le t)  |$, where $k$ and $k'$ are the number of genes in $S$ and $S^c$, respectively. For a given gene set, significance of this test implies that the gene set $S$ is enriched by genes that share similar genetic variants as those for the disease of interest, suggesting their relevance to the etiology of the disease. See Section 4 for their detailed applications.

\section{Simulation Studies}

This section demonstrates and compares the empirical performance of our proposed estimators and some alternative estimators under various  settings. 

\paragraph{Simulation under multivariate Gaussian models.}
We generate a pair of $n$-dimensional vectors, denoted as $\bold{x}_n$ and $\bold{y}_n$ with $ n = 1.5\times 10^5,3\times 10^5$ and $5\times 10^5$, from a multivariate normal distribution $N(\theta,\Sig)$ and $N(\mu,\Sig)$, respectively. We choose $s\in\{50,100,200,400,800\}$, which cover both the regions $s\le \sqrt{n}$ and $s>\sqrt{n}$, and generate $(\theta,\mu)$ as follows: 1) the supports of $\theta$ and $\mu$ are randomly sampled from the coordinates, with the nonzero components generated from Unif(1,10); and  2) we partition the coordinates of $\theta$ and $\mu$ into blocks of size 10 and randomly pick $s/10$ blocks as the support, on which we assign symmetric triangle-shaped values whose maximal value is generated from Unif(5,10). The above signal structures are referred as Sparse Pattern I and II, respectively. For the covariance matrix $\Sig$ we consider a global covariance $\Sig={\bf I}$ and two block-wise covariances $\Sig_1$ and $\Sig_2$ (see Supplementary Material for their explicit forms).
We evaluate the our proposed estimators $\widehat{T}^S_K$, $\widehat{T}_K$ and $\widetilde{T}$, as well as  (1) the hybrid thresholding estimator without sample splitting, denoted as $\widehat{T}^{S*}_K$; and (2) the naive estimator $\overline{T}$ that simply calculates the absolute inner product of observed vectors.
For $\widehat{T}^S_K$ and $\widehat{T}^{S*}_K$, we fix $K=8$, whereas for $\widehat{T}_K$, we set $K=\lfloor \frac{1}{12}\log n\rfloor$. Each setting was repeated 100 times and the performance was evaluated using the empirical version of the rescaled mean square error $\textsc{rmse}(\hat{T}) = \frac{1}{s}\sqrt{{\E(\hat{T}-{T})^2}}$.
Tables \ref{table:t1} reports the empirical \textsc{rmse} of the five estimators under the settings with independent observations. Due to page limit, the results under correlated observations are given in Tables C1 and C2 of the Supplementary Material. In general, the performances of $\widehat{T}^S_K$, $\widetilde{T}$ and $\widehat{T}^S_K$ are roughly the same, with $\widehat{T}^S_K$ having slightly better performance among the three, but all superior to the naive estimator $\overline{T}$. $\widehat{T}^{S*}_K$ outperforms all the other estimators in all the settings, which may due to reduced variability by not using  sample splitting. Since the sample splitting is needed only to facilitate our theoretical analysis, in applications we suggest to use $\widehat{T}^{S*}_K$ for better performance. Moreover, Tables  C1 and C2 in our Supplementary Material shows that the proposed estimators are robust to the underlying sparsity patterns and the covariance structures.

\begin{table}[t!] 
	\caption{Empirical \textsc{rmse}  under covariance $\Sig={\bf I}_n$.   $\widehat{T}^{S*}_K$: the hybrid thresholding estimator without sample splitting;
		$\widehat{T}^S_K$: the hybrid thresholding estimator;
		$\widetilde{T}$: the simple thresholding estimator;
		$\widehat{T}_K$: the hybrid non-thresholding estimator;
		$\overline{T}$: the naive estimator that calculates the absolute inner product of observed vectors.}
	\vskip .2cm
	\centerline{\tabcolsep=3truept\begin{tabular*}{ 0.8 \textwidth}{cc|ccccc|ccccc}	\hline 
			$\frac{n}{10^4}$&   $s$  &  $\widehat{T}^{S*}_K$  & $\widehat{T}^S_K$   & 	$\widetilde{T}$   &      $\widehat{T}_K$  & $\overline{T} $&$\widehat{T}^{S*}_K$  & $\widehat{T}^S_K$   & 	$\widetilde{T}$   &      $\widehat{T}_K$  & $\overline{T}$ \\  
			\hline  
			&&	\multicolumn{5}{c}{Sparse Pattern I} &\multicolumn{5}{c}{Sparse Pattern II}\\
			&50& 10.54 & 20.85 & 27.47 & 25.14 & 1910.3 &8.69 & 26.79 & 32.9 & 28.84 & 1909.2\\
			&100&  11.41 & 21.00 & 27.92 & 25.63 & 954.3&8.08 & 26.33 & 32.64 & 28.75 & 954.3\\
			15  & 200&  10.30 & 21.19 & 30.83 & 28.01 & 476.9&8.42 & 25.83 & 32.33 & 28.54 & 476.9 \\
			& 400  &10.01 & 20.57 & 29.24 & 26.78 & 238.0& 8.64 & 25.88 & 31.67 & 27.84 & 238.0\\
			& 800& 10.58 & 22.36 & 29.99 & 27.05 & 118.8&9.20 & 25.48 & 31.16 & 27.61 & 118.7\\		
			\hline  
			& 50& 9.50 & 20.51 & 30.13 & 27.7 & 3819.4&10.72 & 28.11 & 33.67 & 29.73 & 3819.8\\
			&100&11.07 & 25.85 & 33.66 & 29.98 & 1909.3& 9.20 & 27.90 & 34.36 & 30.04 & 1908.6\\
			30    & 200&  10.60 & 22.19 & 30.3 & 27.09 & 954.4&9.71 & 25.89 & 31.88 & 28.27 & 954.1\\
			& 400  & 10.54 & 22.22 & 30.08 & 26.85 & 476.9&10.73 & 27.79 & 32.3 & 28.61 & 476.7\\
			& 800&  10.86 & 23.52 & 30.62 & 27.24 & 238.2& 8.62 & 26.67 & 34.2 & 30.11 & 238.0 \\
			\hline  
			& 50&12.27 & 27.30 & 32.18 & 28.67 & 6363.4& 12.02 & 25.78 & 27.07 & 24.37 & 6365.3\\
			&100&11.25 & 24.86 & 30.69 & 27.29 & 3182.4&8.54 & 29.67 & 35.99 & 31.4 & 3182.5\\
			50  & 200&  11.02 & 22.48 & 29.39 & 25.88 & 1591.3&9.98 & 29.13 & 34.21 & 29.94 & 1591.3 \\
			& 400  &11.40 & 23.42 & 29.86 & 26.45 & 795.4&12.51 & 25.28 & 28.06 & 25.09 & 795.2 \\
			& 800& 10.85 & 22.85 & 29.40 & 26.11 & 397.2&10.23 & 27.05 & 32.69 & 28.84 & 397.2\\
			\hline
	\end{tabular*}}
	\label{table:t1}
\end{table}

\paragraph{Simulation under model-generated GWAS and eQTL data allowing for population stratification.}
In order to justify our proposed methods for integrative analysis of \textsc{gwas} and e\textsc{qtl} data, we carried out additional numerical experiments  under more realistic settings where the \textsc{gwas}-based genotypes are simualted allowing for population stratification and the corresponding $z$-scores are calculated  from a case-control study that  adjusts for population structure using principal component (PC) scores. Specifically, for the \textsc{gwas} data, we adopted the simulation settings from \cite{astle2009population} where 1000 cases and 1000 controls are drawn from a population of 6000 individuals partitioned into three equal-sized subpopulations. Ancestral minor allele fractions were generated from Unif(0.05,0.5) for all 10,000 unlinked \textsc{snp}s. For each \textsc{snp}, subpopulation allel fractions are generated from the beta-binomial model $\text{Beta}\big(\frac{1-F}{F}p, \frac{1-F}{F}(1-p)\big)$ with population divergence parameter $F=0.1$. We simulate the disease phenotype under a logistic regression model with 20 \textsc{snp} markers each with effect size 0.4. The population disease prevalence is 0.05. To obtain $z$-scores, we fit marginal logistic regression for each \textsc{snp} accounting for the first 2 PCs of the genotypes. For the e\textsc{qtl}  data, 10,000 unlinked \textsc{snp}s are generated independently with minor allele fractions from Unif(0.05,0.5). The gene expression levels of 2000 samples are simulated under a linear regression model with covariates being $s$ \textsc{snp} markers that overlap with the \textsc{gwas} \textsc{snp}s and each has effect size 0.5, and the errors are independently drawn from the standard normal distribution. The e\textsc{qtl}  $z$-scores are obtained from marginal linear regression. The above simulations were repeated for 500 times. The population mean of the $z$-scores corresponding to the truly associated \textsc{snp} markers are approximated by the sample mean of the $z$-scores. 
Table \ref{table:t6} shows the empirical \textsc{rmse}s for the five estimators with $s\in\{5,10,15,20\}$. Again, our proposed estimators $\widehat{T}_K$, $\widehat{T}^S_K$ and $\widetilde{T}$ outperform the naive estimator $\overline{T}$ across all the settings, while $\widehat{T}^{S*}_K$ performs even better. The numerical results agree with our simulations under the multivariate Gaussian settings and suggest the applicability of our proposed methods for integrating \textsc{gwas} and e\textsc{qtl} data.


\begin{table}[t!]
	\centering
	\caption{Empirical \textsc{rmse} for simulated GWAS and eQTL data.
		$\widehat{T}^{S*}_K$: the hybrid thresholding estimator without sample splitting;
		$\widehat{T}^S_K$: the hybrid thresholding estimator;
		$\widetilde{T}$: the simple thresholding estimator;
		$\widehat{T}_K$: the hybrid non-thresholding estimator;
		$\overline{T}$: the naive estimator that calculates the absolute inner product of observed vectors.	}
	\vskip .2cm
	\centerline{\tabcolsep=3truept\begin{tabular*}{ 0.38\textwidth}{c|ccccc}
			\hline
			$s$  &  $\widehat{T}^{S*}_K$  & $\widehat{T}^S_K$   & 	$\widetilde{T}$   &      $\widehat{T}_K$  & $\overline{T} $ \\  
			\hline  
			5& 19.61 & 32.26 & 40.45 & 34.25 & 1318.1\\
			10& 17.42 & 35.27 & 39.87 & 36.80 & 638.9\\
			15& 13.92 & 31.78 & 36.50 & 34.50 & 425.6\\
			20  & 12.77 & 29.18 & 32.72 & 30.52 & 317.7\\
			\hline
	\end{tabular*}}
	\label{table:t6}
\end{table}

\section{Integrative Data Analysis of Human Heart Failure}

Finally, we apply our proposed estimation procedure to identify genes whose expressions are possibly causally linked to heart failure by integrating \textsc{gwas} and e\textsc{qtl} data. \textsc{gwas} results were obtained from a heart failure genetic association study at the University of Pennsylvania, a  prospective study of patients recruited from the University of Pennsylvania, Case Western Reserve University, and the University of Wisconsin, where genotype data were collected from 4,523 controls and 2,206 cases using the Illumina OmniExpress Plus array. \textsc{gwas} summary statistics were calculated controlling for age, gender, and the first two principal components of the genotypes. 

Heart failure e\textsc{qtl} data were obtained from the MAGNet e\textsc{qtl} study (https://www.med.upenn. edu/magnet/index.shtml), where the left ventricular free-wall tissue was collected from 136 donor hearts without heart failure. Genotype data were collected using using Affymetrix genome-wide \textsc{snp} array 6.0 and only markers in Hardy-Weinberg equilibrium with minor allele frequencies above 5\% were considered. Gene expression data were collected using Affymetrix GeneChip ST1.1 arrays, normalized using \textsc{RMA} \citep{irizarry2003exploration} and batch-corrected using ComBat \citep{johnson2007adjusting}. To obtain a common set of  \textsc{snp}s, 
\textsc{snp}s were imputed using 1000 Genomes Project data.  Summary statistics for the MAGNet e\textsc{qtl} data were obtained  using the fast marginal regression algorithm of \citet{sikorska2013gwas} controlling for age and gender.

\subsection{Ranking of potential heart failure causal genes}

\begin{figure}[h!]
	\centering
	\includegraphics[angle=0,width=12cm]{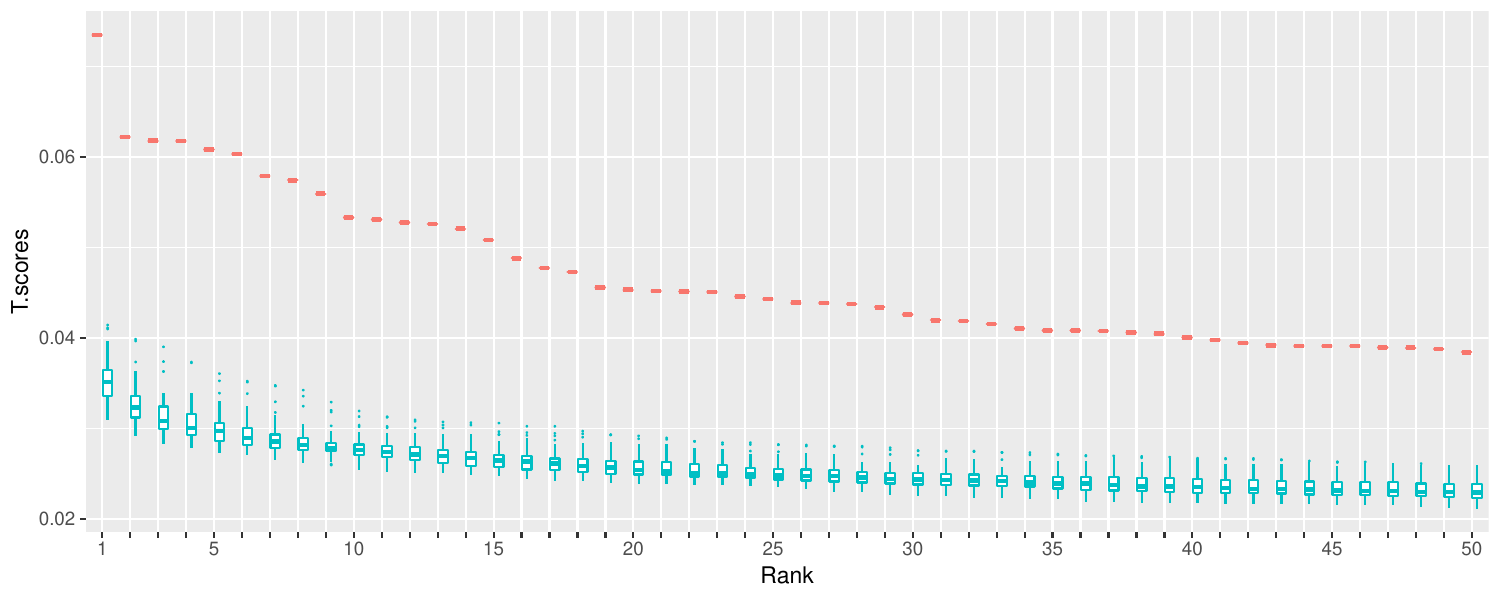}
	\includegraphics[angle=0,width=12cm]{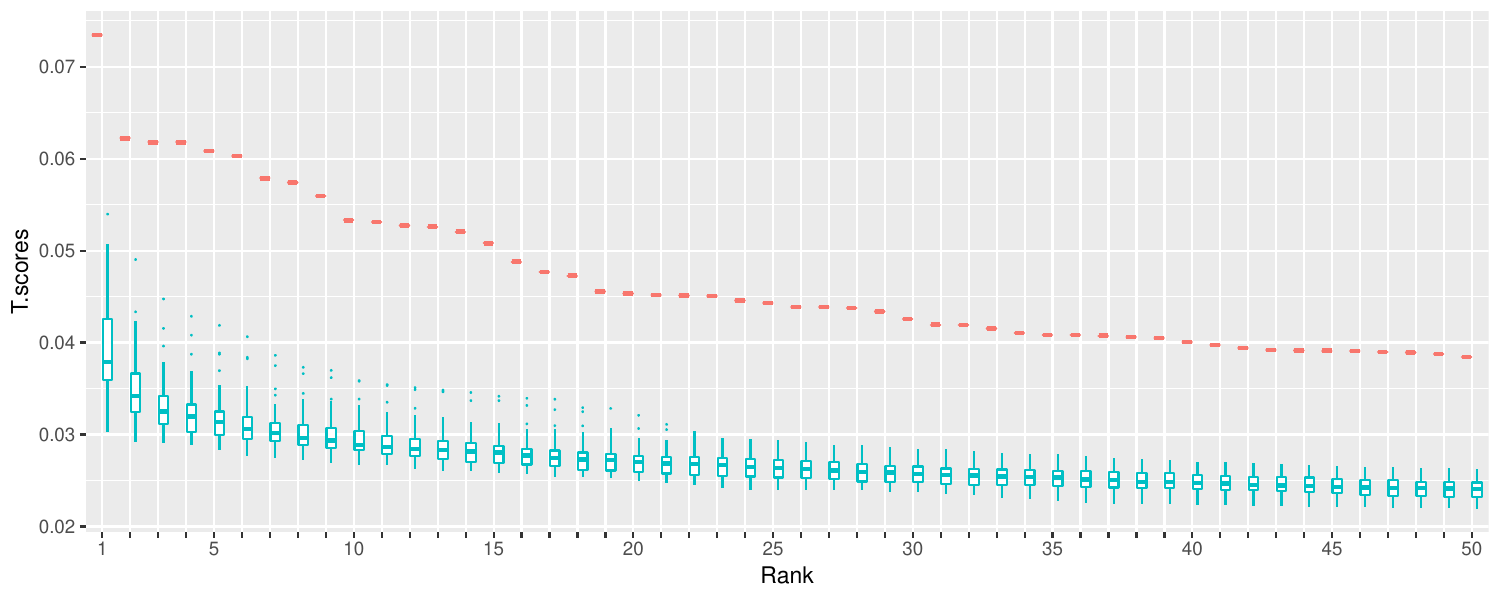}
	\vspace{-0.3cm}
	\caption{Estimated score for the top 50 genes and the box plots of the top scores based on 50 permutations. Top: random permutation of the \textsc{gwas} scores; bottom: cyclic permutations of the \textsc{gwas} scores. }
	\label{permute}
\end{figure}

After matching the  \textsc{snp}s of both e\textsc{qtl} and \textsc{gwas} data, we had  a total of 347,019  \textsc{snp}s and 19,081 genes with expression data available. In light of the results in simulation studies, throughout we use $\widehat{T}^{S*}_K$ with $K=8$ to estimate the T-scores. The analysis then follows from Section 2.5 so that the genes are ordered by their normalized T-scores. 
To assess that the  top scored genes indeed represent true biological signals, we  calculated the $T$-scores for two ``null datasets" that  are created using permutations. For the first dataset, we randomly permuted the labels of the  \textsc{snp}s of the \textsc{gwas} $z$-scores by sampling without replacement before estimating the normalized $T$-scores  with e\textsc{qtl} $z$-scores. For the second dataset, we permuted the labels of the  \textsc{snp}s of the \textsc{gwas} $z$-scores in a circular manner similar to \cite{cabrera2012uncovering}.  Specifically,  for each chromosome, we randomly chose one \textsc{snp} as start of the chromosome and move the \textsc{snp}s  on the fragment  before this \textsc{snp}  to the end.  Such a cyclic permutation preserves the local dependence of the $z$-scores. 
By permuting the data from one phenotype, we break the original matching  of the $z$-scores  between the two phenotypes.  The permutation was performed 50 times and the null distribution of $T$-scores based on the permuted  data was obtained.  Figure \ref{permute} shows the ranked normalized   $T$-scores based on the original data and the box plots of the ranked $z$-scores based on 50 permutations of the $z$-scores.  We find that all the top ranked genes have larger $T$-scores than the ones based on permutations. In addition, about 30 top ranked genes in the top plot  and about 10 top ranked genes in the bottom plot  have true $T$-scores larger than all the $T$-scores from the permuted datasets.  This confirms that the top ranked genes based on their estimated normalized $T$-scores are not due to random noise and indeed represent certain sharing of genetic variants between heart failure and gene expression levels. 

Table \ref{genes}  lists the top eight highest ranked genes along with their biological annotations. All of the genes are either directly or indirectly associated with human heart failure, including 
those related to fibrotic myocardial degeneration, Wnt signalling activity and heart-valve development. It is interesting that our proposed methods can identify these relevant genes using only the gene expression data measured on normal heart tissues. 

\begin{table}
	\caption{Top eight heart failure associated genes based on the estimated normalized $T$-scores and their functional annotations.}\label{genes}
	\vskip .2cm
	\centerline{\tabcolsep=3truept\begin{tabular*}{ 0.8\textwidth}{lll}
			\hline 
			Gene Name &  Annotations  \\  
			\hline  
			TMEM37 &   voltage-gated ion channel activity   \citep{chen2007calcium} \\
			ADCY7 &  adenylate cyclase activity; fibrotic myocardial  \\
			& degeneration \citep{nojiri2006oxidative}\\
			C1QC &  Wnt signaling activity; associated with heart \\
			&  failure \citep{naito2012complement} \\
			FAM98A &  associated with ventricular  septal  defect \citep{liu2018differential}\\
			BMP2 & associated with heart-valve development \\  &\citep{rivera2006bmp2}\\
			SLCO2B1   & organic anion transporter; associated with cardiac glycoside \\
			&  \citep{mikkaichi2004organic} \\
			C1QA  & Wnt signaling activity; associated with heart \\
			&  failure \citep{naito2012complement} \\
			FCGR2B &  intracellular signaling activity; associated with vascular \\
			& disease pathogenesis \citep{tanigaki2015fcgamma}\\
			\hline
	\end{tabular*}}
	\label{table:t4}
\end{table}

\subsection{Gene set enrichment analysis}
To complete our analysis, we finish this section with the gene set enrichment analysis (\textsc{gsea}) \citep{subramanian2005gene} using the normalized $T$-scores to identify the heart failure associated biological processes.    In the following analysis, we removed genes with low expression and small variability across the samples, which resulted in a total of 6,355 genes. 
The method described in Section 2.5 was applied to the gene sets from Gene Ontology (\textsc{go}) (\citealt{botstein2000gene}) that contain at least 10 genes, and 5,023 biological processes were tested.
Figure C1 in our Supplementary Material  presents the directed acyclic graphs of the \textsc{go} biological processes  that linked to  the most significant \textsc{go} terms  from the simultaneous signal \textsc{gsea} analysis.  Table \ref{table:t5} shows the top 6 \textsc{go} biological processes identified from the \textsc{gsea} analysis. Among them, regulation of skeletal muscle contraction, linoleic acid metabolic process and calcium ion regulation are strongly implicated in  human heart failure. \cite{murphy2011cardiovascular} showed that skeletal muscle reflexes are essential to the initiation and regulation of the cardiovascular response to exercise, and alteration of this reflex mechanism can happen in disease states such as hypertension and heart failure. In \cite{farvid2014dietary}, a thorough meta-analysis was carried out, which supports a significant inverse association between dietary linoleic acid intake, when replacing either carbohydrates or saturated fat, and risk of coronary heart disease. Moreover,  the importance of calcium-dependent signaling in the heart failure was reported in  \cite{marks2003calcium}, who suggested that impaired calcium release causes decreased muscle contraction (systolic dysfunction) and defective calcium removal hampers relaxation (diastolic dysfunction).

\begin{table}
	\centering
	\caption{Top six \textsc{go} biological processes that are associated with heart failure based on the gene set enrichment analysis}
	\vspace{0.3cm}
	\begin{tabular}{lc}
		\hline 
		\textsc{go} term & $p$-value  \\  
		\hline  
		\emph{Biological Process} &\\
		regulation of skeletal muscle contraction by regulation of release&\\
		of sequestered calcium ion & $7.9\times 10^{-7}$\\
		linoleic acid metabolic process & $1.0\times 10^{-6}$\\
		regulation of skeletal muscle contraction by calcium ion signaling & $3.4\times 10^{-6}$ \\
		positive regulation of sequestering of calcium ion & $3.4\times 10^{-6}$\\
		cellular response to caffeine& $1.0\times 10^{-5}$\\
		cellular response to purine-containing compound & $1.0\times 10^{-5}$\\
		\hline
	\end{tabular}
	\label{table:t5}
\end{table}

\section{Discussion}

This paper considers the optimal estimation over sparse parameter spaces. However, in Section 2, the minimax rates of convergence were established for the parameter spaces $D^{\infty}(n^{\beta},L_n)$ with $\beta\in (0,1/2)\cup (1/2,1)$, leaving a gap at $\beta=1/2$. Our theoretical analysis suggests a lower bound  (\ref{sparse.lower.equation4}) with the rate function $\psi(s,n)\asymp  1$, which cannot be attained by any of our proposed estimators. Nevertheless, in Section B.1 of our Supplementary Material, we confirm that $L_n^2s^2/n^2$ is the minimax rate of convergence for $\beta=1/2$ by proposing an estimator achieving such  rate.

In some applications, we may need to consider non-sparse parameter spaces. In this case, our theoretical analysis shows that the estimator $\widehat{T}_K$ with $K=r\log n$ for some small constant $r>0$ can still be applied. Specifically, from our proof of Theorem 1 and Theorem 2, it follows that, if we define the non-sparse parameter space as $\mathcal{D}^{\infty}_U(L_n)=\big\{(\theta,\mu,\Sig_1,\Sig_2):  (\theta,\mu)\in \R^n\times \R^n,\max(\|\theta\|_\infty , \|\mu\|_\infty )\le L_n,  \Sig_1=\Sig_2={\bf I}_n \big\}$ with $L_n\gtrsim \sqrt{\log n}$, then for $\bold{x_n} \sim N(\theta,\Sig_1)$ and $\bold{y_n} \sim N(\mu,\Sig_2)$, the minimax rate $
\inf_{\hat{T}}\sup_{(\theta,\mu,\Sig_1,\Sig_2)\in \mathcal{D}^{\infty}_U(L_n) } \mathcal{R}(\hat{T} ) \asymp \frac{L_n}{ \log n}$
can be attained by the above $\widehat{T}_K$.

In light of our genetic applications, it is also natural and interesting to consider parameter spaces where $\theta$ and $\mu$ are both sparse in themselves. Specifically, assuming triple sparsity of $\theta$, $\mu$ and $\{\theta_i\mu_i\}_{i=1}^n$, interesting phase transitions might exist, where the minimax rates and the optimal estimators could be different from those reported in the current paper. 
In addition to the estimation problems, it is also of interest to conduct hypothesis testing or construct confidence intervals for $T$-score. These problems can be technically challenging due to the non-smooth functional. We leave these important problems for future investigations.

\vskip 14pt
\noindent {\large\bf Supplementary Materials}

Our supplementary material includes proofs of the main theorems. Some supplementary notes, figures and tables are also included.
\par
\vskip 14pt
\noindent {\large\bf Acknowledgements}

The authors are grateful to the Editor, the Associate Editor and two anonymous referees for
their comments and suggestions that improve the presentation of the paper. R.M. would like to thank Mark G. Low for helpful discussions. The research reported in this publication was supported by NIH grants R01GM123056 and R01GM129781 and NSF grant DMS-1712735.
\par

\bibliographystyle{chicago}
\bibliography{reference}

\newpage

\title{Supplement to ``Optimal Estimation of Simultaneous Signals Using Absolute Inner Product with Applications to Integrative Genomics"}
\author{Rong Ma$^1$, T. Tony Cai$^2$ and Hongzhe Li$^1$ \\
	Department of Biostatistics, Epidemiology and Informatics$^1$\\
	Department of Statistics$^2$\\
	University of Pennsylvania\\
	Philadelphia, PA 19104}
\date{}
\maketitle
\thispagestyle{empty}

\begin{abstract}
This supplementary material includes proofs of the main theorems and the technical lemmas of the paper ``Optimal Estimation of Simultaneous Signals Using Absolute Inner Product with Applications to Integrative Genomics." Some supplementary notes, figures and tables are also included.
\end{abstract}

\setcounter{section}{0}

\section{Proofs of the Minimax Lower Bounds}

\subsection{Proof of Theorem 1}

By the definition of the rate function $\psi(s,n)$, it suffices to show the following two statements.
\begin{equation} \label{sparse.lower.equation1.pf}
\inf_{\hat{T}} \sup_{\substack{(\theta,\mu,\Sig_1,\Sig_2) \in D^{\infty}(s,L_n)}} \mathbb{E}( \hat{T} -T(\theta,\mu))^2   \gtrsim L_n^2s^2\cdot \min\bigg\{ \log\bigg(1+\frac{n}{s^2}\bigg),L^2_n\bigg\},
\end{equation}
\begin{equation} \label{sparse.lower.equation2.pf}
\inf_{\hat{T}} \sup_{\substack{(\theta,\mu,\Sig_1,\Sig_2) \in D^{\infty}(s,L_n)}} \mathbb{E}( \hat{T}-T(\theta,\mu))^2  \gtrsim \frac{L_n^2s^2}{\log^2 s} \cdot \min\{ \log s, L_n^2  \}.
\end{equation}
\paragraph{Proof of (\ref{sparse.lower.equation1.pf}).}
Let $l(s,n)$ be the class of all subsets of $\{1,\dots,n \}$ of $s$ elements. For $I \in l(s,n)$, we denote $\theta_{I}  = \{ \theta \in \R^n : \theta_i =0, \forall i \notin I,\text{and } \theta_i=\rho, \forall i\in I  \}$. 
Suppose that $\mu$ is fixed with $\mu=\mu^*$ where $\mu_i^*=L_n$ for all $1\le i\le n$. Denote
\[
g_{I} (x_1,\dots, x_n,y_1,\dots, y_n) = \Pi_{i=1}^n \psi_{\theta_{I,i}} (x_i)\Pi_{i=1}^n \psi_{\mu^*_i} (y_i),
\]
where $\psi_{\theta_i}$ denotes the density function of $N(\theta_i,1)$ and  $\theta_{I,i}$ is the $i$-th component of $\theta_I$. In this way, we are considering the class of probability measures for $\{x_i,y_i\}_{1\le i\le n}$ where the mean vector for $\{x_i\}$ is the $s$-sparse vector $\theta_I$ whereas the mean vector for $\{y_i\}$ is $\mu^*$. By averaging over all the possible $I\in l(s,n)$, we have the mixture probability measure
\[
g = \frac{1}{ {n \choose s}} \sum_{I \in l(s,n)} g_I.
\]
On the other hand, we consider the probability measure
\[
f = \Pi_{i=1}^n \phi(x_i)\Pi_{i=1}^n  \psi_{\mu^*_i} (y_i)
\]
where $\phi$ is the normal density of $N(0,1)$. From the above construction, we consider $D^{\theta}_s(\rho)=\{(\theta,\mu,\Sig_1,\Sig_2): \theta=\theta_I, I\in l(s,n), \mu=\mu^*, \Sig_1=\Sig_2=I\}\cup \{(\theta,\mu,\Sig_1,\Sig_2): \theta=0, \mu=\mu^*, \Sig_1=\Sig_2=I\}$. Apparently, for $\rho\le L_n$, $D^{\theta}_s(\rho) \subset D^{\infty}(s,L_n)$.  In the following, we will consider the $\chi^2$-divergence between $g$ and $f$ and obtain the minimax lower bound over $D_s(\rho)$ using the constrained risk inequality obtained by \cite{brown1996constrained}.
Note that
\[
\int \frac{g^2}{f} = \frac{1}{{n \choose s}^2} \sum_{I \in l(s,n)} \sum_{I' \in l(s,n)} \int \frac{g_{I} g_{I'}}{f}
\]
and for any $I$ and $I'$,
\begin{align*}
\int\frac{g_{I} g_{I'}}{f}&=\frac{1}{(2\pi)^{n/2}} \int \exp\bigg\{ -\frac{\sum_i(x_i -\theta_{I,i})^2+\sum_i(x_i-\theta_{I',i} )^2 -\sum_ix_i^2}{2}  \bigg\} \\
&\quad\times \frac{1}{(2\pi)^{n/2}} \int \exp\bigg\{ -\frac{\sum_i(y_i -\mu^*_{i})^2}{2}  \bigg\} \\
&= \frac{1}{(2\pi)^{n/2}} \int \exp\bigg\{ -\frac{\sum_i(x_i - \theta_{I,i}-\theta_{I',i})^2 - 2\sum_{i=1}^n\theta_{I,i}\theta_{I',i} }{2}   \bigg\} \\
& = \exp(\rho^2 j)
\end{align*}
where $j$ is the number of points in the set $I \cap I'$. It follows that
\begin{align*}
\int \frac{g^2}{f} &= \frac{1}{{n \choose s}^2} \sum_{j=1}^{s} {n \choose s} {s\choose j}{n-s \choose s-j} \exp(2\rho^2 j)  \\
& =\E \exp(2\rho^2J)
\end{align*}
where $J$ has a hypergeometric distribution
\[
\mathbb{P} (J=j) = \frac{{s \choose j}{n-s \choose s-j}}{{n \choose s}}.
\]
As shown in p.173 of \citep{aldous1985exchangeability}, $J$ has the same distribution as the random variable $\E(Z| \mathcal{B}_n)$ where $Z$ is a binomial random variable of parameters $(s,s/n)$ and $\mathcal{B}_n$ some suitable $\sigma$-algebra. Thus, by Jensen's inequality we have
\begin{equation}
\E \exp(2J \rho^2) \le \bigg(1-\frac{s}{n}+\frac{s}{n}e^{2\rho^2}\bigg)^{s}.
\end{equation}
The rest of the proof will be separated into two parts, corresponding to $L_n\ge\sqrt{\log \big(1+\frac{n}{s^2}\big)}$ and $L_n< \sqrt{\log \big(1+\frac{n}{s^2}\big)}$, respectively. 

\underline{Case I. $L_n\ge \sqrt{\log \big(1+\frac{n}{s^2}\big)}$.} 
By taking $\rho  = \sqrt{\log \big(1+\frac{n}{s^2}\big)}\le L_n$,  we have 
\begin{equation}
\int \frac{g^2}{f} =\E \exp(J \rho^2) \le e.
\end{equation}
Then if some estimator $\delta$ satisfies
\begin{equation}
\E_f(\delta-0)^2 \le C s^2\|\mu^*\|_\infty^2\log\bigg(1+\frac{n}{s^2}\bigg)
\end{equation}
then by the constrained risk inequality (Theorem 1 of  \cite{brown1996constrained}),
\begin{align*}
\E_g\bigg(\delta-s\rho\|\mu^*\|_\infty \bigg)^2 &\ge s^2\rho^2\|\mu^*\|_{\infty}^2 - 2\rho s\|\mu^*\|_{\infty} C^{1/2}s\|\mu^*\|_{\infty} \log^{1/2} \bigg(1+\frac{n}{s^2}\bigg) \\
& ={s^2\|\mu^*\|_{\infty}^2} \log \bigg(1+\frac{n}{s^2}\bigg) - \sqrt{2C}s^2\|\mu^*\|_{\infty}^2\log \bigg(1+\frac{n}{s^2}\bigg),
\end{align*}
for any such estimator $\delta$.
Recall that $\|\mu^*\|_{\infty}=L_n$. By choosing $ C$ sufficiently small, we conclude that there exists some $I \in l(s,n)$ such that
\begin{equation}
\E_{g_I}\bigg(\delta -s\rho\|\mu^*\|_\infty \bigg)^2 \ge Cs^2L_n^2\log \bigg(1+\frac{n}{s^2}\bigg)
\end{equation}
for all $\delta$. Therefore we have
\begin{align} \label{bound4}
\inf_{\hat{T}} \sup_{(\theta,\mu,\Sig_1,\Sig_2)\in D^{\theta}_s(\rho)} \mathbb{E}( \hat{T} -T(\theta,\mu))^2 &\ge   Cs^2L_n^2 \log \bigg(1+\frac{n}{s^2}\bigg).
\end{align}
The lower bound (\ref{sparse.lower.equation1.pf}) then follows from the fact that $L_n\gtrsim \sqrt{\log n}$.

\underline{Case II. $L_n<\sqrt{\log \big(1+\frac{n}{s^2}\big)}$.} 
By taking $\rho  =  L_n<\sqrt{\log \big(1+\frac{n}{s^2}\big)}$,  again we have 
\begin{equation}
\int \frac{g^2}{f} =\E \exp(J \rho^2) \le e.
\end{equation}
Then if some estimator $\delta$ satisfies
\begin{equation}
\E_f(\delta-0)^2 \le C s^2\|\mu^*\|_\infty^2L_n^2
\end{equation}
then by the constrained risk inequality (Theorem 1 of  \cite{brown1996constrained}),
\begin{align*}
\E_g\bigg(\delta-s\rho\|\mu^*\|_\infty \bigg)^2 &\ge s^2\rho^2\|\mu^*\|_{\infty}^2 - 2\rho s\|\mu^*\|_{\infty} C^{1/2}s\|\mu^*\|_{\infty} L_n \\
& ={s^2\|\mu^*\|_{\infty}^2} L_n^2 - \sqrt{2C}s^2\|\mu^*\|_{\infty}^2L_n^2,
\end{align*}
for any such estimator $\delta$.
Recall that $\|\mu^*\|_{\infty}=L_n$. By choosing $ C$ sufficiently small, we conclude that there exists some $I \in l(s,n)$ such that
\begin{equation}
\E_{g_I}\bigg(\delta -s\rho\|\mu^*\|_\infty \bigg)^2 \ge Cs^2L_n^4
\end{equation}
for all $\delta$. Therefore we have
\begin{align} \label{bound5}
\inf_{\hat{T}} \sup_{(\theta,\mu,\Sig_1,\Sig_2)\in D^{\theta}_s(\rho)} \mathbb{E}( \hat{T} -T(\theta,\mu))^2 &\ge   Cs^2L_n^4.
\end{align}
This proves the other part of (\ref{sparse.lower.equation1.pf}).

\paragraph{Proof of (\ref{sparse.lower.equation2.pf}).}
Now we prove the second part of the theorem.
It follows by Lemma 1 in \cite{cai2011testing} that there exist measures $\nu_{i}$ on $[-M_n,M_n]$ for $i=0, 1$, such that:
\begin{enumerate}
	\item $\nu_{0}$ and $\nu_{1}$ are symmetric around 0;
	\item $\int t^l \nu_{1}(dt) =\int t^l \nu_{0}(dt)  $, for $l=0,1,\dots, k_n$;
	\item $\int |t| \nu_{1}(dt) - \int |t| \nu_{0}(dt) = 2M_n\delta_{k_n}  $.
	\item { $\int |t| \nu_{0}(dt) >0$.}
\end{enumerate}
where $\delta_{k_n}$ is the distance in the uniform norm on $[-1,1]$ from the absolute value function $f(x) = |x|$ to the space of polynomials of no more than degree $k_n$. In addition, $\delta_{k_n} = \beta_*k_n^{-1}(1+o(1))$ as $k_n \rightarrow \infty$.
Now we consider product priors on the $n$-vector $\theta$, which are supported on the first $s\le n$ components. Let 
\[
\nu_{i1}^n = \Pi^{\otimes s}\nu_{i}\cdot \Pi^{\otimes n-s}1_{\{0\}} ,\quad\quad \nu_{i2}^n = \Pi^{\otimes n}1_{\{\mu^*\}}
\]
for $i=0$ and 1. In other words, we put independent priors $\nu_{i}$ for the first $s$ components of the vector $\theta$, while keeping the other coordinates as $0$, and we fix $\mu=\mu^*$.

Following the above construction, we have
\[
\E_{\nu_{11}^n}\frac{1}{n}\sum_{i=1}^n|\theta_i| - \E_{\nu_{01}^n}\frac{1}{n}\sum_{i=1}^n|\theta_i|= \frac{s}{n}\bigg[\E_{\nu_{11}}|\theta_i|-\E_{\nu_{01}}|\theta_i| \bigg]= 2sM_n\delta_{k_n}/n,
\] 
and
\[
\E_{\nu_{12}^n}\frac{1}{n}\sum_{i=1}^n|\mu_i| - \E_{\nu_{02}^n}\frac{1}{n}\sum_{i=1}^n|\mu_i|=0
\] 
Then we have
\begin{align}
\E_{\nu_{11}^n\nu_{12}^n}\frac{1}{n}T(\theta,\mu)-\E_{\nu_{01}^n\nu_{02}^n}\frac{1}{n}T(\theta,\mu) &=\frac{s}{n}\big(\E_{\nu_{11}}|\theta_i| \E_{\nu_{12}}|\mu_i| - \E_{\nu_{01}}|\theta_i| \E_{\nu_{02}}|\mu_i| \big)\nonumber \\
& = \frac{s}{n}(\E_{\nu_{11}}|\theta_i| \E_{\nu_{12}}|\mu_i| - \E_{\nu_{01}}|\theta_i|\E_{\nu_{12}}|\mu_i|)\nonumber\\
&\quad + \frac{s}{n} (\E_{\nu_{01}}|\theta_i|\E_{\nu_{12}}|\mu_i|- \E_{\nu_{01}}|\theta_i| \E_{\nu_{02}}|\mu_i| )\nonumber \\
& =\E_{\nu_{12}}|\mu_i| \frac{s}{n}(\E_{\nu_{11}}|\theta_i|-\E_{\nu_{01}}|\theta_i|)+\E_{\nu_{01}}|\theta_i|\frac{s}{n}(\E_{\nu_{12}}|\mu_i|- \E_{\nu_{02}}|\mu_i|) \nonumber \\
&=\frac{2sM_n\delta_{k_n}}{n} \E_{\nu_{12}}|\mu_i|  \nonumber \\
& = \frac{2sM_nL_n\delta^2_{k_n}}{n} 
\end{align}
We further have
\beq
V_0^2\equiv \frac{1}{n^2}\E_{\nu_{01}^n\nu_{02}^n}(T(\theta,\mu)-\E_{\nu_{01}^n\nu_{02}^n}T(\theta,\mu))^2 \le \frac{sM_n^2L_n^2}{n^2}
\eeq
Set $f_{0,M_n}(y) = \int\phi(y-t)\nu_0(dt)$ and $f_{1,M_n}(y) = \int\phi(y-t)\nu_1(dt)$. Note that since $g(x) = \exp(-x)$ is a convex function of $x$, and $\nu_0$ is symmetric,
\begin{align*}
f_{0,M_n}(y) &\ge \frac{1}{\sqrt{2\pi}} \exp\bigg( -\int\frac{(y-t)^2}{2}\nu_0(dt)   \bigg)\\
& = \phi(y)\exp\bigg( -\frac{1}{2}M_n^2\int t^2 v'_0(dt) \bigg)\\
& \ge \phi(y)\exp \bigg( -\frac{1}{2}M_n^2 \bigg).
\end{align*}
Let $H_r$ be the Hermite polynomial defined by
\begin{equation}
\frac{d^r}{dy^r}\phi(y) = (-1)^rH_r(y)\phi(y)
\end{equation}
which satisfiy
\begin{equation}
\int H^2_r(y)\phi(y) dy = r! \quad\quad\text{and}\quad\quad \int H_r(y)H_l(y)\phi(y) dy =0
\end{equation}
when $r\ne l$. Then
\[
\phi(y-t) = \sum_{k=0}^\infty H_k(y)\phi(y)\frac{t^k}{k!}
\]
and it follows that
\begin{align*}
\int\frac{(f_{1,M_n}(y) - f_{0,M_n}(y))^2}{f_{0,M_n}(y)} dy &\le \int  (f_{1,M_n}(y) - f_{0,M_n}(y))^2 e^{M_n^2/2}/\phi(y) dy \\
&= e^{M_n^2/2} \int \bigg\{  \sum_{k=0}^\infty H_k(y) \frac{\phi(y)}{k!}\bigg[ \int t^k \nu_1(dt) - \int t^k \nu_0(dt)\bigg]     \bigg\}^2 \bigg/\phi(y)  dy \\
& = e^{M_n^2/2} \int \sum_{k=k_n+1}^\infty H^2_k(y)\phi(y) \frac{M_n^{2k}}{(k!)^2}\bigg[\int t^k \nu_1(dt) - \int t^k \nu_0(dt) \bigg]^2 dy \\
& = e^{M_n^2/2} \sum_{k=k_n+1}^\infty \frac{M_n^{2k}}{k!} \bigg[\int t^k \nu_1(dt) - \int t^k \nu_0(dt) \bigg]^2 \\
& \le e^{M_n^2/2} \sum_{k=k_n+1}^{\infty}\frac{M_n^{2k}}{k!}.
\end{align*}
It then follows
\begin{align*}
I_n^2 &=  \prod_{i=1}^s \int \frac{( f_{1,M_n}(x_i))^2}{ f_{0,M_n}(x_i)} dx_i - 1 \le \bigg( 1+e^{M_n^2/2} \sum_{k=k_n+1}^\infty \frac{1}{k!} M_n^{2k} \bigg)^s-1 \le \bigg( 1+e^{M_n^2/2} D\frac{1}{k_n!} M_n^{2k_n} \bigg)^s-1,
\end{align*}
for some $D>0$.
Since $k! > (k/e)^k$, we also have
\beq
I_n^2 \le  \bigg( 1+e^{M_n^2/2} D\bigg( \frac{ eM_n^2}{k_n}\bigg)^{k_n} \bigg)^s-1
\eeq
The rest of the proof is separated into two parts.

\underline{Case I. $L_n\ge \sqrt{\log s}$.}
Now let $M_n= \sqrt{\log s}\le L_n$ and $k_n\asymp \log s $. It then can be checked that $I_n <c$ for some sufficiently small constant $c>0$. 
Therefore, Corollary 1 in \cite{cai2011testing} along with the fact that $L_n \ge \sqrt{\log s}$ yields
\begin{align}
\inf_{\hat{T}}\sup_{(\theta,\mu,\Sig_1,\Sig_2)\in D^{\infty}(s,L_n)} \frac{1}{n^2}\E(\hat{T}-T(\theta,\mu))^2
&\ge \frac{(2sM_n\delta_{k_n}L_n/n- (s^{1/2}M_nL_n/n)I_n)^2}{(I_n+2)^2}  \nonumber \\
& \ge  \frac{Cs^2L_n^2}{ n^2\log s}.
\end{align}

\underline{Case II. $L_n<\sqrt{\log s}$.}
In this case, we set $M_n= L_n<\sqrt{\log s}$ and $k_n\asymp \log s $. Then again $I_n <c$ for some sufficiently small constant $c>0$. 
Therefore, Corollary 1 in \cite{cai2011testing} along with the fact that $M_n\le L_n$ yields
\begin{align}
\inf_{\hat{T}}\sup_{(\theta,\mu,\Sig_1,\Sig_2)\in D^{\infty}(s,L_n)} \frac{1}{n^2}\E(\hat{T}-T(\theta,\mu))^2
&\ge \frac{(2sM_n\delta_{k_n}L_n/n- (s^{1/2}M_nL_n/n)I_n)^2}{(I_n+2)^2}  \nonumber \\
& \ge  \frac{Cs^2L_n^4}{ n^2\log^2 s}.
\end{align}

\section{Proofs of the Risk Upper Bounds}

\subsection{Proof of Theorem 2}

For the hybrid estimators defined in Section 2.2 of the main paper, the key is to study the bias and variance of a single component. Let $x_1, x_2 \sim N(\theta,1)$, $y_1, y_2 \sim N(\mu,1)$. Denote ${\delta}(x) = \min\{ S_K(x), n^2\}$.  In the following, we analyse the two hybrid estimators separately.

\paragraph{Part I: Analysis of $\widehat{T}_K^S$.} Let
\begin{align}
\xi = \xi(x_1,x_2, y_1,y_2) &=  [{\delta}(x_{1})I(\sqrt{2\log n}<|x_{2}|\le 2\sqrt{2\log n})+|x_{1}|I(|x_{2}|> 2\sqrt{2\log n})]\nonumber \\
&\quad \times  [ {\delta}(y_{1})I(\sqrt{2\log n}<|y_{2}|\le 2\sqrt{2\log n})+|y_{1}|I(|y_{2}|> 2\sqrt{2\log n})] 
\end{align}
Note that 
\begin{align}
\E(\xi) &= [\E{\delta}(x_{1})P(\sqrt{2\log n}<|x_{2}|\le 2\sqrt{2\log n})+\E|x_{1}|P(|x_{2}|> 2\sqrt{2\log n})]\nonumber \\
&\quad \times  [ \E{\delta}(y_{1})P(\sqrt{2\log n}<|y_{2}|\le 2\sqrt{2\log n})+\E|y_{1}|P(|y_{2}|> 2\sqrt{2\log n})] 
\end{align}
We denote
\[
\tilde{\sigma}^2_x = \V( {\delta}(x_{1})I(\sqrt{2\log n}<|x_{2}|\le 2\sqrt{2\log n})+|x_{1}|I(|x_{2}|> 2\sqrt{2\log n})),
\]
\[
\tilde{\sigma}^2_y = \V({\delta}(y_{1})I(\sqrt{2\log n}<|y_{2}|\le 2\sqrt{2\log n})+|y_{1}|I(|y_{2}|> 2\sqrt{2\log n})),
\]
and
\[
\tilde{\theta}_x = \E{\delta}(x_{1})P(\sqrt{2\log n}<|x_{2}|\le 2\sqrt{2\log n})+\E|x_{1}|P(|x_{2}|> 2\sqrt{2\log n}),
\]
\[
\tilde{\mu}_y = \E{\delta}(y_{1})P(\sqrt{2\log n}<|y_{2}|\le 2\sqrt{2\log n})+\E|y_{1}|P(|y_{2}|> 2\sqrt{2\log n}).
\]
Then we have
\[
\V(\xi) = \tilde{\sigma}_x^2 \tilde{\sigma}_y^2+ \tilde{\sigma}_x^2 \tilde{\mu}_y^2+ \tilde{\sigma}_y^2 \tilde{\theta}_x^2
\]
The following two propositions are key to our calculation of the estimation risk.

\begin{proposition}
	For all $\theta,\mu \in \R$, we have
	\[
	|B_x|  \equiv|\tilde{\theta}_x - |\theta||\lesssim \sqrt{\log n}, \quad\quad  |B_y | \equiv|\tilde{\mu}_y - |\mu| | \lesssim \sqrt{\log n},
	\]
	and
	\[
	\tilde{\sigma}^2_x \lesssim \log n , \quad\quad \tilde{\sigma}^2_y \lesssim \log n .
	\]
	In particular, when $\theta= 0$, we have $|B_x| \le n^{-2}{\log n}$ and $\tilde{\sigma}^2_x \lesssim n^{-1}\log n $, whereas when $\mu=0$, we have $|B_y | \le n^{-2}{ \log n},$ and $ \tilde{\sigma}^2_y \lesssim n^{-1}\log n$.
\end{proposition}

\begin{proposition}
	For all $\theta,\mu \in \R$ such that $\theta,\mu\le L_n$ where $L_n\le(\sqrt{2}-1)\sqrt{\log n}$, we have
	\[
	|B_x|  \equiv|\tilde{\theta}_x - |\theta||\lesssim L_n, \quad\quad  |B_y | \equiv|\tilde{\mu}_y - |\mu| | \lesssim L_n,
	\]
	and
	\[
	\tilde{\sigma}^2_x \lesssim \frac{\log n}{\sqrt{n}}, \quad\quad \tilde{\sigma}^2_y \lesssim \frac{\log n}{\sqrt{n}}.
	\]
\end{proposition}

Now the bias of the estimator $\xi$ satisfies
\begin{align} \label{bias20}
B(\xi) &= \E\xi - |\theta||\mu| = \tilde{\theta}_x\tilde{\mu}_y - |\theta||\mu| \nonumber  \\
& \le |\theta| \cdot |\tilde{\mu}_y - |\mu| | + |\mu| \cdot |\tilde{\mu}_x - |\theta|| +|\tilde{\mu}_y - |\mu| |\cdot|\tilde{\mu}_x - |\theta||\nonumber \\
&\le |\mu||B_y|+ |\theta||B_x| + |B_x||B_y|.
\end{align}
and the variance 
\begin{align} \label{var20}
\V(\xi) &= \tilde{\sigma}^2_x\tilde{\sigma}^2_y +\tilde{\sigma}^2_x\tilde{\mu}_y^2 + \tilde{\sigma}^2_y\tilde{\theta}_x^2 \nonumber \\
&\lesssim \tilde{\sigma}^2_x\tilde{\sigma}^2_y +  |\mu|^2 \tilde{\sigma}^2_x +|{\theta}|^2 \tilde{\sigma}^2_y+{B_x^2 \tilde{\sigma}^2_y}+B_y^2 \tilde{\sigma}^2_x.
\end{align}
Now  let $(x_{1\ell},x_{2\ell},...,x_{n\ell})\sim N(\theta,{\bf I}_n)$ and $(y_{1\ell},y_{2\ell},...,y_{n\ell})\sim N(\mu,{\bf I}_n)$ for $\ell=1,2$, and let 
\[
\widehat{T}_K^S =  \sum_{i=1}^n \xi (x_{i1},x_{i2}, y_{i1},y_{i2}).
\]
\underline{Case I. $L\gtrsim \sqrt{\log n}$.}
It follows from (\ref{bias20}) and Cauchy-Schwartz inequality that the bias of $\widehat{T}_K^S$ is bounded by
\[
|B(\widehat{T}_K^S)| \lesssim (\|\theta\|_\infty+\|\mu\|_\infty) s\sqrt{\log n}+s\log n
\]
From (\ref{var20}) we have the variance of $\widehat{T^S(\theta,\mu)}$ is bounded by
\begin{align}
\V( \widehat{T}_K^S) &\le  \sum_{i=1}^n \V(\xi (x_{i1},x_{i2}, y_{i1},y_{i2})) \nonumber \\
& \lesssim {s \log^2 n} + (\|\theta\|_{\infty}^2+\|\mu\|_{\infty}^2)s\log n
\end{align}
Therefore the mean squared error of $\widehat{T}_K^S$ satisfies
\begin{align*}
\E(\widehat{T}_K^S-T(\theta,\mu))^2 &\le B^2(\widehat{T}_K^S) + \V( \widehat{T}_K^S)  \\
& \lesssim s^2L_n^2\log n.
\end{align*}

\underline{Case II. $L_n\lesssim \sqrt{\log n}$.} It follows that
\[
|B(\widehat{T}_K^S)| \lesssim (\|\theta\|_\infty+\|\mu\|_\infty) sL_n+sL_n^2
\]
From (\ref{var20}) we have the variance of $\widehat{T^S(\theta,\mu)}$ is bounded by
\begin{align}
\V( \widehat{T}_K^S) &\le  \sum_{i=1}^n \V(\xi (x_{i1},x_{i2}, y_{i1},y_{i2})) \nonumber \\
& \lesssim \log^2 n+\frac{sL_n^2\log n}{\sqrt{n}}+\frac{\log^2 n}{\sqrt{n}}+L_n^2\log n+\frac{\log^3 n}{n^3}.
\end{align}
Therefore the mean squared error of $\widehat{T}_K^S$ satisfies
\begin{align*}
\E(\widehat{T}_K^S-T(\theta,\mu))^2 &\le B^2(\widehat{T}_K^S) + \V( \widehat{T}_K^S)  \\
& \lesssim s^2L_n^4+\frac{\log^2 n}{\sqrt{n}}+L_n^2\log n
\end{align*}

\paragraph{Part II: Analysis of $\widehat{T}_K$.}
With a slight abuse of notation, we denote
\begin{align}
\xi = \xi(x_1,x_2, y_1,y_2) &=  [{\delta}(x_{1})I(|x_{2}|\le 2\sqrt{2\log n})+|x_{1}|I(|x_{2}|> 2\sqrt{2\log n})]\nonumber \\
&\quad \times  [ {\delta}(y_{1})I(|y_{2}|\le 2\sqrt{2\log n})+|y_{1}|I(|y_{2}|> 2\sqrt{2\log n})] 
\end{align}
Note that 
\begin{align}
\E(\xi) &= [\E{\delta}(x_{1})P(|x_{2}|\le 2\sqrt{2\log n})+\E|x_{1}|P(|x_{2}|> 2\sqrt{2\log n})]\nonumber \\
&\quad \times  [ \E{\delta}(y_{1})P(|y_{2}|\le 2\sqrt{2\log n})+\E|y_{1}|P(|y_{2}|> 2\sqrt{2\log n})] 
\end{align}
We denote
\[
\tilde{\sigma}^2_x = \V( {\delta}(x_{1})I(|x_{2}|\le 2\sqrt{2\log n})+|x_{1}|I(|x_{2}|> 2\sqrt{2\log n})),
\]
\[
\tilde{\sigma}^2_y = \V({\delta}(y_{1})I(|y_{2}|\le 2\sqrt{2\log n})+|y_{1}|I(|y_{2}|> 2\sqrt{2\log n})),
\]
and
\[
\tilde{\theta}_x = \E{\delta}(x_{1})P(|x_{2}|\le 2\sqrt{2\log n})+\E|x_{1}|P(|x_{2}|> 2\sqrt{2\log n}),
\]
\[
\tilde{\mu}_y = \E{\delta}(y_{1})P(|y_{2}|\le 2\sqrt{2\log n})+\E|y_{1}|P(|y_{2}|> 2\sqrt{2\log n}).
\]

\begin{proposition}
	Let $B_x= \tilde{\theta}_x-|\theta|$, $B_y= \tilde{\mu}_y-|\mu|$. For all $\theta,\mu \in \R$ and $K=r \log n$ for some $0<r<1/3$, we have
	\[
	|B_x| \lesssim \frac{1}{\sqrt{\log n}}, \quad\quad |B_y |\lesssim \frac{1}{\sqrt{\log n}},
	\]
	and
	\[
	\tilde{\sigma}^2_x =O\big( n^{6r}\log^3 n\big)  , \quad\quad \tilde{\sigma}^2_y = O\big( n^{6r}\log^3 n\big).
	\]
	In particular, when $\theta= 0$, we have $|B_x| \le n^{6r-2}\log^3 n$ and $\tilde{\sigma}^2_x \lesssim n^{6r}\log^3 n $, whereas when $\mu=0$, we have $|B_y | \le n^{6r-2}\log^3 n,$ and $ \tilde{\sigma}^2_y \lesssim n^{6r}\log^3 n$.
\end{proposition}
Again, let $(x_{1\ell},x_{2\ell},...,x_{n\ell})\sim N(\theta,{\bf I}_n)$ and $(y_{1\ell},y_{2\ell},...,y_{n\ell})\sim N(\mu,{\bf I}_n)$ for $\ell=1,2$, and let $\widehat{T}_K=  \sum_{i=1}^n \xi (x_{i1},x_{i2}, y_{i1},y_{i2})$.
It follows that,  when $0<r<1/4$,  the bias can be bounded by
\[
|B(\widehat{T}_K)| \lesssim \frac{\|\theta\|_{\infty} s}{\sqrt{\log n}} +\frac{\|\mu\|_\infty s}{\sqrt{\log n}} + \frac{s}{\log n}.
\]
On the other hand, the variance of $\widehat{T(\theta,\mu)}$ is bounded by
\begin{align}
\V( \widehat{T}_K) &\le  \sum_{i=1}^n \V(\xi (x_{i1},x_{i2}, y_{i1},y_{i2})) \nonumber \\
& \lesssim n^{12r+1}\log^6n + n^{6r+1}\log^3 n \cdot{(\|\theta\|_{\infty}^2+\|\mu\|_{\infty}^2)}\\
&\lesssim |B(\widehat{T}_K)|^2,
\end{align}
as long as $0<r<\frac{2\beta-1}{12}$. In this case, we have
\begin{align*}
\E(\widehat{T}_K-T(\theta,\mu))^2 
& \lesssim \frac{s^2}{\log^2n}+\frac{s^2\|\theta\|_{\infty}^2}{\log n}+\frac{s^2\|\mu\|_{\infty}^2}{\log n}.
\end{align*}
The final result follows from the fact that $\max(\|\theta\|_\infty,\|\mu\|_\infty)\le L_n$.

\subsection{Proof of Theorem 3}

Let $x_1, x_2 \sim N(\theta,1)$ and $y_1, y_2 \sim N(\mu,1)$. Define
\begin{align}
\xi = \xi(x_1,x_2, y_1,y_2) &=  [|x_{1}|I(|x_{2}|> 2\sqrt{2\log n})] \times  [|y_{1}|I(|y_{2}|> 2\sqrt{2\log n})]. 
\end{align}
Note that 
\begin{align}
\E(\xi) &= \E|x_{1}|P(|x_{2}|> 2\sqrt{2\log n})\times   \E|y_{1}|P(|y_{2}|> 2\sqrt{2\log n})
\end{align}
Denote
\[
\tilde{\sigma}^2_x = \V(|x_{1}|I(|x_{2}|> 2\sqrt{2\log n})),\quad\quad
\tilde{\sigma}^2_y = \V(|y_{1}|I(|y_{2}|> 2\sqrt{2\log n})),
\]
and
\[
\tilde{\theta}_x = \E|x_{1}|P(|x_{2}|> 2\sqrt{2\log n}),
\quad\quad
\tilde{\mu}_y = \E|y_{1}|P(|y_{2}|> 2\sqrt{2\log n}).
\]
Then we have
\[
\V(\xi) = \tilde{\sigma}_x^2 \tilde{\sigma}_y^2+ \tilde{\sigma}_x^2 \tilde{\mu}_y^2+ \tilde{\sigma}_y^2 \tilde{\theta}_x^2
\]
\begin{proposition}
	Let $B_x = \tilde{\theta}_x-|\theta|$ and $B_y = \tilde{\mu}_y-|\mu|$. For all $\theta,\mu \in \R$, we have 
	\[
	|B_x| \lesssim \sqrt{\log n}, \quad\quad |B_y | \lesssim \sqrt{\log n},
	\]
	and
	\[
	\tilde{\sigma}^2_x \lesssim \log n , \quad\quad \tilde{\sigma}^2_y \lesssim \log n .
	\]
	In particular, when $\theta=0$ we have $|B_x| \lesssim n^{-4}$ and $\tilde{\sigma}^2_x\lesssim n^{-4}$, whereas when $\mu=0$ we have $|B_y| \lesssim n^{-4}$ and $\tilde{\sigma}^2_y\lesssim n^{-4}$.
\end{proposition}

\begin{proposition}
	For all $\theta,\mu \in \R$ such that $\mu,\theta\le L_n$ where $L_n\le \sqrt{2\log n}$, we have 
	\[
	|B_x| \lesssim L_n, \quad\quad |B_y | \lesssim L_n,
	\]
	and
	\[
	\tilde{\sigma}^2_x \lesssim \frac{\log n}{n} , \quad\quad \tilde{\sigma}^2_y \lesssim \frac{\log n}{n}.
	\]
\end{proposition}

Now the bias of the estimator $\xi$ is
\begin{align} \label{bias2}
B(\xi) &= \E\xi - |\theta||\mu| = \tilde{\theta}_x\tilde{\mu}_y - |\theta||\mu| \nonumber  \\
& \le |\theta| |B_y|+ |\mu| |B_x| +|B_x||B_y|,
\end{align}
whereas the variance is bounded by
\begin{align} \label{var2}
\V(\xi) = \tilde{\sigma}^2_x\tilde{\sigma}^2_y +\tilde{\sigma}^2_x\tilde{\mu}_y^2 + \tilde{\sigma}^2_y\tilde{\theta}_x^2 \lesssim \tilde{\sigma}^2_x\tilde{\sigma}^2_y +  |\mu|^2 \tilde{\sigma}^2_x +|{\theta}|^2 \tilde{\sigma}^2_y+\log n (\tilde{\sigma}^2_y+ \tilde{\sigma}^2_x).
\end{align}
Now let $(x_{1\ell},x_{2\ell},...,x_{n\ell})\sim N(\theta,{\bf I}_n)$ and $(y_{1\ell},y_{2\ell},...,y_{n\ell})\sim N(\mu,{\bf I}_n)$ for $\ell=1,2$, and let 
\[
\widetilde{T}=  \sum_{i=1}^n \xi (x_{i1},x_{i2}, y_{i1},y_{i2}).
\]

\underline{Case I. $L_n\ge \sqrt{2\log n}$.}
It then follows  that
\begin{align*}
|B(\widetilde{T})| &\le s\log n+(\|\theta\|_\infty+\|\mu\|_\infty)s\sqrt{\log n},
\end{align*}
and 
\begin{align}
\V( \widetilde{T}) &\le s\log^2n+(\|\theta\|_\infty^2+\|\mu\|_\infty^2)s\log n.
\end{align}
Therefore the mean squared error of $\widetilde{T}$ satisfies
\begin{align*}
\E(\widetilde{T}-T(\theta,\mu))^2 &\le B^2(\widetilde{T}) + \V(\widetilde{T})  \\
& \lesssim s^2\log n(\|\theta\|_{\infty}^2+\|\mu\|_{\infty}^2+\log n).
\end{align*}
The final result follows from the fact that $\max(\|\theta\|_\infty,\|\mu\|_\infty)\le L_n$.

\underline{Case II. $L_n< \sqrt{2\log n}$.} It then follows  that
\begin{align*}
|B(\widetilde{T})| &\le sL_n^2+(\|\theta\|_\infty+\|\mu\|_\infty)sL_n,
\end{align*}
and 
\begin{align}
\V( \widetilde{T}) &\le \frac{\log^2 n}{n}+{L_n^2\log n}.
\end{align}
Therefore the mean squared error of $\widetilde{T}$ satisfies
\begin{align*}
\E(\widetilde{T}-T(\theta,\mu))^2 &\le B^2(\widetilde{T}) + \V(\widetilde{T})  \\
& \lesssim s^2L_n^4+\frac{\log^2 n}{n}+{L_n^2\log n}.
\end{align*}

\subsection{Proof of Theorem 4}

By our sample splitting argument, it suffices to obtain the mean squared risk bound for the estimator $\widetilde{T}=\sum_{i=1}^n \hat{U}_i(x_i)\hat{U}_i(y_i)$ of $T(\theta,\mu)=\sum_{i=1}^n|\theta_i||\mu_i|$ where $\hat{U}_i(x_i)=|x_{i1}|I(|x_{i2}|>2\sqrt{2\log n})$, $\hat{U}_i(y_i)=|y_{i1}|I(|y_{i2}|>2\sqrt{2\log n})$ and $(x_{1\ell},x_{2\ell},...,x_{n\ell})\sim N(\theta,\Sig_1)$ and $(y_{1\ell},y_{2\ell},...,y_{n\ell})\sim N(\mu,\Sig_2)$ for $\ell=1,2$. By Proposition 3, the bias of the estimator $\xi_i=\hat{U}_i(x_i)\hat{U}_i(y_i)$ satisfies
\begin{align} \label{bias2.cor}
B(\xi_i) = \E\xi_i - |\theta_i||\mu_i| \le  |\mu_i||B_{ix}|+|\theta_i||B_{iy}| +|B_{ix}||B_{iy}|,
\end{align}
where $B_{ix}=\hat{U}_i(x_i)-|\theta_i|$ and $B_{iy}=\hat{U}_i(y_i)-|\mu_i|$.
The variance satisfies
\begin{align} \label{var2.cor}
\V(\xi_i) &\lesssim \tilde{\sigma}^2_{ix}\tilde{\sigma}^2_{iy} +  |\mu_i|^2 \tilde{\sigma}^2_{ix} +|{\theta}_i|^2 \tilde{\sigma}^2_{iy}+\log n (\tilde{\sigma}^2_{iy}+ \tilde{\sigma}^2_{ix}),
\end{align}
where $ \tilde{\sigma}^2_{ix}=\V(\hat{U}_i(x_i))$ and $ \tilde{\sigma}^2_{iy}=\V(\hat{U}_i(y_i))$.
The covariance between two copies
\begin{align*}
&\quad\text{Cov}(\xi_i,\xi_j)=\E \xi_i\xi_j-\E\xi_i\E\xi_j\\
&=\E \hat{U}_i(x_i)\hat{U}_i(y_i)\hat{U}_j(x_j)\hat{U}_j(y_j)-\E \hat{U}_i(x_i)\hat{U}_i(y_i)\E\hat{U}_j(x_j)\hat{U}_j(y_j)\\
&=\E \hat{U}_i(x_i)\hat{U}_j(x_j)\E\hat{U}_i(y_i)\hat{U}_j(y_j)-\E \hat{U}_i(x_i)\E\hat{U}_j(x_j)\E\hat{U}_i(y_i)\E\hat{U}_j(y_j)
\end{align*}
Thus, we have
\begin{align*}
|\text{Cov}(\xi_i,\xi_j)| &\le |\text{Cov}(\hat{U}_i(x_i),\hat{U}_j(x_j))|\cdot \tilde{\mu}_i\tilde{\mu}_j+|\text{Cov}(\hat{U}_i(y_i),\hat{U}_j(y_j))|\cdot \tilde{\theta}_i\tilde{\theta}_j\\
&\quad+|\text{Cov}(\hat{U}_i(x_i),\hat{U}_j(x_j))\text{Cov}(\hat{U}_i(y_i),\hat{U}_j(y_j))|,
\end{align*}
where $\tilde{\theta}_j=\E\hat{U}_j(x_j)$ and $\tilde{\mu}_j=\E\hat{U}_j(y_j)$.
Note that
\begin{align*}
&\quad|\text{Cov}(\hat{U}_i(x_i),\hat{U}_j(x_j))|\le |\E\hat{U}_i(x_i)\hat{U}_j(x_j)|+|\tilde{\theta}_{i}\tilde{\theta}_{j}|
\end{align*}
where
\begin{align*}
&\E\hat{U}_i(x_i)\hat{U}_j(x_j)=\E|x_{i2}||x_{j2}|P(|x_{i2}|> 2\sqrt{2\log n},|x_{j2}|> 2\sqrt{2\log n}).
\end{align*}
Suppose one of $\theta_i$ and $\theta_j$ is $0$, and the other bounded by $L_n$. Then by the proof of Proposition 3, we have
\[
\E\hat{U}_i(x_i)\hat{U}_j(x_j)=O(n^{-4}L_n^2)
\]
and
\[
|\tilde{\theta}_{i}\tilde{\theta}_{j}|=O(n^{-4}L_n).
\]
So 
\[
|\text{Cov}(\hat{U}_i(x_i),\hat{U}_j(x_j))|=O(n^{-4}L_n^2).
\]
As a result, since $\mu_i,\mu_j\lesssim L_n$, we have
\[
|\text{Cov}(\xi_i,\xi_j)|\le O(n^{-4}L_n^4).
\]
Hence, summation over $O(n^2)$ terms will be bounded by $ O(n^{-2}L_n^4)$. On the other hand, if neither $\theta_i$ or $\theta_j$ is zero, we have the trivial bound from Proposition 3
\[
|\text{Cov}(\xi_i,\xi_j)|\le \log^2 n,
\]
and the summation over $O(s^2)$ terms will be bounded by $s^2\log^2n$. Thus, as long as $L_n\lesssim \sqrt{n}$, we have
\begin{align}
\V( \widetilde{T}) &\le \sum_{i=1}^n \V(\xi (x_{i1},x_{i2}, y_{i1},y_{i2})) + O(n^{-2}L_n^4)+O(s^2\log^2n) \lesssim {s^2\log^2 n }.
\end{align}
Now note that $|B(\widetilde{T})| \lesssim s\log n+(\|\theta\|_\infty+\|\mu\|_\infty)s\sqrt{\log n}$, it follows that
\begin{align*}
\E(\widetilde{T}-T(\theta,\mu))^2 \lesssim {s^2\log^2 n}+L_n^2s^2\log n.
\end{align*}

\section{Proof of Propositions 1-5}

\subsection{Proof of Proposition 1}

In the following, we divide into four cases according to the value of $|\theta|$. When $|\theta| = 0$, we show that we are actually estimating $|\theta|$ by 0. When $0\le |\theta| \le \sqrt{2\log n}$, we show that the estimator $\xi'$ behaves essentially like ${\delta}$, which is a good estimators when $|\theta|$ is small. When $\sqrt{2\log n} <|\theta| \le 4\sqrt{2\log n}$, we show that the hybrid estimator $\xi$ uses either $\delta(x_1)$ or $|x_1|$ and in this case both are good estimators of $| \theta|$. When $|\theta|$ is large, the hybrid estimator is essentially the same as $|x_1|$. We need the following lemmas to facilitate our proof.

\bel \label{constant.lem}
Consider $G_K(x)$ defined in the main paper. The constant term of $\tilde{G}_K(x) = \sum_{l=0}^K \tilde{g}_{2l}x^{2l}$, with $\tilde{g}_{2l} = M_n^{-2l+1}g_{2l}$, satisfies
\begin{equation} \label{3}
\tilde{g}_0=M_ng_{0} \le \frac{2M_n}{\pi(2K+1)}.
\end{equation}
\eel

\bel \label{moments.s.lem}
Let $X\sim N(\theta,1)$ and ${S}_K(x) = \sum_{k=1}^K g_{2k}M_n^{-2k+1}H_{2k}(x)$. Then for all $|\theta| \le 4\sqrt{2 \log n}$, we have
\[
\bigg|\E {S}_K(X) -{|\theta|} \bigg| \le \frac{4M_n}{\pi(2K+1)},
\]
and for $M_n^2\ge K$, we have $\E {S}^2_K(X) \lesssim 2^{8K}M_n^2 K^2.$
\eel

\bel
\label{var.lem2}
Suppose $I(A)$ and $I(B)$ are indicator random variables independent of $X$ and $Y$, with $A\cap B = \emptyset$ then
\begin{align} \label{var21}
\V(X I(A)+YI(B)) &= \V(X)P(A)+\V(Y)P(B)+(\E X)^2P(A)P(A^c) \nonumber \\
&\quad+(\E Y)^2P(B)P(B^c)-2\E X\E Y P(A)P(B).
\end{align}
In particular, if $A^c = B$, then we have
\begin{equation} \label{var22}
\V(X I(A)+YI(A^c)) = \V(X)P(A)+\V(Y)P(A^c)+(\E X-\E Y)^2P(A)P(A^c).
\end{equation}
\eel
Applying Lemma \ref{var.lem2}, we have
\begin{align*}
\tilde{\sigma}^2_x &= \V({\delta}(x_1)) {P}(\sqrt{2\log n}<|x_2|\le 2\sqrt{2 \log n}) + \V(|x_1|) {P}(|x_2|>2\sqrt{2 \log n}) \\
&\quad +(\E {\delta}(x_1))^2{P}(\sqrt{2\log n}<|x_2|\le 2\sqrt{2 \log n}) (1-{P}(\sqrt{2\log n}<|x_2|\le 2\sqrt{2 \log n}))\\
&\quad +(\E |x_1|)^2{P}(|x_2|>2\sqrt{2 \log n})(1-{P}(|x_2|>2\sqrt{2 \log n}))\\
&\quad-2\E {\delta}(x_1)\E |x_1|{P}(\sqrt{2\log n}<|x_2|\le 2\sqrt{2 \log n}){P}(|x_2|>2\sqrt{2 \log n}).
\end{align*}
\begin{align*}
\tilde{\sigma}^2_y &= \V({\delta}(y_1)) {P}(\sqrt{2\log n}<|x_2|\le 2\sqrt{2 \log n}) + \V(|y_1|) {P}(|y_2|>2\sqrt{2 \log n}) \\
&\quad +(\E {\delta}(y_1))^2{P}(\sqrt{2\log n}<|y_2|\le 2\sqrt{2 \log n}) (1-{P}(\sqrt{2\log n}<|y_2|\le 2\sqrt{2 \log n}))\\
&\quad +(\E |y_1|)^2{P}(|y_2|>2\sqrt{2 \log n})(1-{P}(|y_2|>2\sqrt{2 \log n}))\\
&\quad-2\E {\delta}(y_1)\E |y_1|{P}(\sqrt{2\log n}<|y_2|\le 2\sqrt{2 \log n}){P}(|y_2|>2\sqrt{2 \log n}).
\end{align*}

\paragraph{Case 1.} $\theta = 0$. Note that ${\delta}(x_1)$ can be written as
\[
{\delta}(x_1) = {S}_K(x_1) -({S}_K(x_1)-n^2)I({S}_K(x_1)\ge n^2).
\]
Consequently,
\begin{align*}
| B_x | &= \big| \E({\delta}(x_1)) {P}(\sqrt{2\log n} <|x_2|\le 2\sqrt{2 \log n}) + \E(|x_1|) {P}(|x_2|>2\sqrt{2 \log n}) \big| \\
& \le |  \E {S}_K(x_1) | +\E \{({S}_K(x_1)-n^2)I({S}_K(x_1)\ge n^2)\}  +  \E |x_1|  {P}(|x_2|>2\sqrt{2 \log n}) \\
& \equiv B_1+B_2+B_3.
\end{align*}
By definition of  ${S}_K(x_1)$ we have
\begin{equation}
B_1 = 0.
\end{equation}
It follows from the standard bound for normal tail probability $\Phi(-z) \le z^{-1} \phi(z)$ for $z>0$ that
\begin{equation}
{P}(|x_2| >2\sqrt{2\log n}) = 2\Phi(-2\sqrt{2\log n}) \le \frac{1}{2\sqrt{\pi\log n}} n^{-4}.
\end{equation}
And in this case 
\begin{equation}
\E |x_1| = 2\phi(0).
\end{equation}
It then follows that
\begin{equation}
B_3 \le  2\phi(0) \cdot \frac{1}{2\sqrt{\pi\log n}} n^{-4} = \frac{1}{\pi \sqrt{2\log n}}n^{-4}.
\end{equation}
Now consider $B_2$. Note that for any random variable $X$ and any constant $\lambda >0$,
\[
\E (XI(X\ge \lambda))\le \lambda^{-1} \E(X^2 I(X\ge\lambda) \le \lambda^{-1} \E X^2.
\]
This together with Lemma \ref{moments.s.lem} yields that
\begin{equation} \label{b2.s}
B_2\le \E \{({S}_K(x_1)I({S}_K(x_1)\ge n^2)\} \le n^{-2} \E({S}^2_K(x_1)) \lesssim n^{-2}\log n.
\end{equation}
Combining the three pieces together, we have
\[
|B_x| \le B_1+B_2+B_3 \lesssim \frac{\log n}{ n^{2}}.
\]
We now consider the variance. It follows that
\begin{align*}
\tilde{\sigma}^2_x &\le\V({S}_K(x_1)){P}(\sqrt{2\log n} <|x_2|\le 2\sqrt{2 \log n}) + \V(|x_1|) {P}(|x_2|>2\sqrt{2 \log n}) \\
&\quad +(\E {\delta}(x_1))^2 +(\E |x_1|)^2{P}(|x_2|>2\sqrt{2 \log n})-2\E {\delta}(x_1)\E |x_1|{P}(|x_2|>2\sqrt{2 \log n})\\
&\le {\E}{S}^2_K(x_1)n^{-1}+(\E {\delta}(x_1))^2+[{\E}x_1^2+({\E}|x_1|)^2 -2\E {\delta}(x_1)\E |x_1|] \cdot \frac{1}{2 n^4 \sqrt{\pi\log n}}\\
& \lesssim n^{-1}2^{8K}M_n^2 K^2 + n^{-4}\log^4 n+\frac{1}{n^4\sqrt{\log n}}\\
& \lesssim n^{-1}\log n
\end{align*}
where we use the fact that
\[
{P}(\sqrt{2\log n} <|x_2|\le 2\sqrt{2 \log n})  \le {P}(\sqrt{2\log n}\le |x|) \le \Phi(-\sqrt{2\log n}) \le n^{-1}.
\]

\paragraph{Case 2.} $0< | \theta| \le\sqrt{2\log n}$. In this case
\begin{align*}
| B_x | &= | \E({\delta}(x_1)) {P}(\sqrt{2\log n} <|x_2|\le 2\sqrt{2 \log n}) + \E(|x_1|) {P}(|x_2|>2\sqrt{2 \log n}) -| \theta || \\
&\le |  \E{S}_K(x_1)-| \theta| | + \E \{({S}_K(x_1)-n^2)I({S}_K(x_1)\ge n^2)\}  \\
&\quad + ( \E |x_1|  ){P}(|x_2|>2\sqrt{2 \log n}) +|\theta|(1-{P}(\sqrt{2\log n} <|x_2|\le 2\sqrt{2 \log n}) )\\
& \equiv B_1+B_2+B_3+B_4
\end{align*}
From Lemma \ref{moments.s.lem} we have
\[
B_1 \lesssim \sqrt{\log n}.
\]
Again, the standard bound for normal tail probability yields
\[
{P}(|x_2| > 2\sqrt{2\log n}) \le 2\Phi(-\sqrt{2\log n}) \le \frac{1}{\sqrt{\pi \log n}}n^{-1}
\]
Note that
\[
\E |x_1| = |\theta| +2\phi(\theta) - 2|\theta|\Phi(-|\theta|)\le|\theta| +1\le\sqrt{2\log n}+1.
\]
Then we have
\[
B_3 \le \bigg(\sqrt{2\log n}+1  \bigg)\cdot \frac{1}{\sqrt{\pi \log n}}n^{-1} \le 3 n^{-1},
\]
and
\[
B_4 \le |\theta| \le\sqrt{2\log n}.
\]
Note that $B_2$ follows (\ref{b2.s}), and we have
\[
|B_x | \le B_1+B_2+B_3+B_4 \lesssim \sqrt{\log n}.
\]
For the variance, note that
\begin{align*}
(\E {\delta}(x_1))^2 &\le  \E {\delta}^2(x_1)  \\
& = \E ( \min \{{S}^2_K(x_1), n^4\}) \\
& =  \E \big[  {S}^2_K(x_1) - ( {S}^2_K(x_1)-n^4)I( {S}^2_K(x_1)>n^4)  \big]\\
& \le \E {S}^2_K(x_1) \\
& \lesssim  \log n,
\end{align*}
and 
\begin{align*}
{\E}x_1^2+({\E}|x_1|)^2 -2 \E{\delta}(x_1)\E |x_1|  &\le \V(x_1)+(\E |x_1|)^2 +(\sqrt{2\log n})^2 \lesssim \log n
\end{align*}
Then we have
\begin{align*}
\tilde{\sigma}^2_x &\le \V({S}_K(x_1))+ \V(|x_1|){P}(|x_2|>2\sqrt{2 \log n}) +(\E {\delta}(x_1))^2 \\
&\quad +(\E |x_1|)^2{P}(|x_2|>2\sqrt{2 \log n})+2|\E{\delta}(x_1)|\cdot \E |x_1|{P}(|x_2|>2\sqrt{2 \log n})\\
&\le {\E}{S}^2_K(x_1)+(\E {\delta}(x_1))^2+[{\E}x_1^2+({\E}|x_1|)^2 +2|\E {\delta}(x_1)|\cdot \E |x_1|] \cdot \frac{1}{ n \sqrt{\pi\log n}}\\
& \le 2 \log n+\frac{5 \sqrt{\log n}}{\sqrt{\pi}n}\\
& \lesssim \log n.
\end{align*}

\paragraph{Case 3} $\sqrt{2\log n} \le |\theta| \le 4\sqrt{2\log n}$. In this case, 
\begin{align*}
|B_x| &= | \E({\delta}(x_1)) {P}(\sqrt{2\log n} <|x_2|\le 2\sqrt{2 \log n}) + \E(|x_1|) {P}(|x_2|>2\sqrt{2 \log n}) -| \theta || \\
&\le | \E ({\delta}(x_1)) -| \theta| |+ |\E |x_1| -|\theta| | +|\theta| {P}(|x_2|\le 2\sqrt{2\log n})\\
&\le | \E{S}_K(x_1)-| \theta| | + \E \{({S}_K(x_1)-n^2)I({S}_K(x_1)\ge n^2)\}  + 2\phi(\theta)+4\sqrt{2\log n}\\
& \lesssim \sqrt{\log n} + n^{-2}\log n + n^{-1}\\
& \lesssim \sqrt{\log n}.
\end{align*}
For the variance, similarly since
\begin{align*}
{\E}x_1^2+({\E}|x_1|)^2 +2|\E{\delta}(x_1)|\cdot \E |x_1|  &\le \V(x_1)+(\E |x_1|)^2 +(4\sqrt{2\log n})^2\\
&\quad+\bigg(4\sqrt{2\log n}+\frac{2 M_n}{\pi K} \bigg) 4\sqrt{2\log n}\\
& \le 1 + (32+32+32) \log n +\frac{64\sqrt{2} \log n}{\pi K}\\
& \lesssim  \log n,
\end{align*}
and then
\begin{align*}
\tilde{\sigma}^2_x &\le \V({S}_K(x_1))+ \V(|x_1|) +( \E {\delta}(x_1))^2 +(\E |x_1|)^2+2|\E {\delta}(x_1)|\cdot \E |x_1|\\
&\le  {\E}{S}^2_K(x_1)+( \E {\delta}(x_1))^2+[{\E}x_1^2+({\E}|x_1|)^2 +2\E{\delta}(x_1)\E |x_1|] \\
& \lesssim \log n.
\end{align*}

\paragraph{Case 4.} $|\theta| >4\sqrt{2\log n}$. In this case, the standard bound for normal tail probability yields that
\[
{P}(|x_2| \le 2\sqrt{2\log n}) \le 2\Phi(-(|\theta|/-2\sqrt{2\log n})) \le 2\Phi\bigg( -\frac{|\theta|}{2} \bigg) \le \frac{4}{|\theta|}\phi\bigg( \frac{|\theta|}{2} \bigg).
\]
In particular, 
\[
{P}(|x_2| \le 2\sqrt{2\log n})  \le 2\Phi(-2\sqrt{2\log n}) \le \frac{1}{2\sqrt{\pi \log n}}n^{-4}.
\]
Also note that
\begin{align*}
\E {\delta}(x_1) & =  \E \min\{ {S}_K(x_1), n^2\}\\
& =  \E ({S}_K(x_1)1\{ {S}_K(x_1)\le n^2 \} +n^2 1\{ {S}_K(x_1)>n^2\} )\\
&\le  n^2 
\end{align*}
Hence,
\begin{align*}
|B_x| &\le | \E({\delta}(x_1)) {P}(\sqrt{2\log n} <|x_2|\le 2\sqrt{2 \log n}) + \E(|x_1|) {P}(|x_2|>2\sqrt{2 \log n}) -| \theta || \\
&\le |\E ({\delta}(x_1)) |{P}(\sqrt{2\log n} <|x_2|\le 2\sqrt{2 \log n})+ |\E |x_1| -|\theta| | +\E|x_1|  {P}(|x_2|\le 2\sqrt{2\log n})\\
&\le |\E |x_1| -|\theta| | + (|\E {\delta}(x_1)| +\E|x_1| ){P}(|x_2|\le2\sqrt{2 \log n}) \\
&\le 2 \phi(\theta) +( n^2+ |\theta|+1) {P}(|x_2|\le2\sqrt{2 \log n}) \\
& \le 2 \phi(\theta) + 4\phi\bigg(\frac{|\theta|}{2}\bigg) +\frac{1}{2}n^{-2}\\
&\le 6\phi\bigg(\frac{\theta}{2}\bigg)+\frac{1}{2n^2}\\
& \le \frac{1}{n^2}.
\end{align*}
For the variance, similarly we have
\begin{align*}
\tilde{\sigma}^2_x &\le \V({\delta}(x_1)) {P}(\sqrt{2\log n}<|x_2|\le 2\sqrt{2 \log n}) + \V(|x_1|)+(\E |x_1|)^2{P}(|x_2|\le2\sqrt{2 \log n})  \\
&\quad +(\E {\delta}(x_1))^2{P}(\sqrt{2\log n}<|x_2|\le 2\sqrt{2 \log n}) \\
&\quad+2|\E {\delta}(x_1)|\cdot \E |x_1|{P}(\sqrt{2\log n}<|x_2|\le 2\sqrt{2 \log n})\\
&\le 1 + [ \V({\delta}(x_1))+( \E {\delta}(x_1))^2+(\E |x_1|)^2+2|\E {\delta}(x_1)|\cdot E |x_1|] {P}(|x_2|\le2\sqrt{2 \log n})\\
& \le 1+[\log n + ( n^2 +|\theta|+1)^2]{P}(|x_2|\le2\sqrt{2 \log n})\\
& = 1+o(1).
\end{align*}
Obviously, the same argument holds for $y_1$, $y_2$ and $|\mu|$. 
\qed

\subsection{Proof of Proposition 2}

When $|\theta|\le L_n \le (\sqrt{2}-1)\sqrt{\log n}$, we have
\begin{align*}
| B_x | &= | \E({\delta}(x_1)) {P}(\sqrt{2\log n} <|x_2|\le 2\sqrt{2 \log n}) + \E(|x_1|) {P}(|x_2|>2\sqrt{2 \log n}) -| \theta || \\
&\le |  \E{S}_K(x_1)-| \theta| |{P}(\sqrt{2\log n} <|x_2|\le 2\sqrt{2 \log n}) + \E \{({S}_K(x_1)-n^2)I({S}_K(x_1)\ge n^2)\}  \\
&\quad + ( \E |x_1|  ){P}(|x_2|>2\sqrt{2 \log n}) +|\theta|(1-{P}(\sqrt{2\log n} <|x_2|\le 2\sqrt{2 \log n}) )\\
& \equiv B_1+B_2+B_3+B_4
\end{align*}
Note that
\begin{align*}
{P}(\sqrt{2\log n} <|x_2|\le 2\sqrt{2 \log n}) &\le{P}(\sqrt{2\log n} <|x_2|)\\
&\le  {P}(\sqrt{2\log n} -L_n<|z|)\\
&\le  2\Phi(-\sqrt{2\log n}+L_n) \\
&\lesssim \frac{1}{n^{1/2}}.
\end{align*}
From Lemma \ref{moments.s.lem} we have
\[
B_1 \lesssim \sqrt{\log n}{P}(\sqrt{2\log n} <|x_2|\le 2\sqrt{2 \log n}) \le \sqrt{\frac{\log n}{n}}.
\]
Again, the standard bound for normal tail probability yields
\[
{P}(|x_2| > 2\sqrt{2\log n}) \le 2\Phi(-\sqrt{2\log n}) \le \frac{1}{\sqrt{\pi \log n}}n^{-1}
\]
Note that
\[
\E |x_1| = |\theta| +2\phi(\theta) - 2|\theta|\Phi(-|\theta|)\le|\theta| +1\le L_n+1.
\]
Then we have
\[
B_3 \le \bigg(\sqrt{2\log n}+1  \bigg)\cdot \frac{1}{\sqrt{\pi \log n}}n^{-1} \le 3 n^{-1},
\]
and
\[
B_4 \le |\theta| \le L_n.
\]
Note that $B_2$ follows (\ref{b2.s}), and we have
\[
|B_x | \le B_1+B_2+B_3+B_4 \lesssim L_n.
\]
For the variance, note that
\begin{align*}
(\E {\delta}(x_1))^2 &\le  \E {\delta}^2(x_1)  \\
& = \E ( \min \{{S}^2_K(x_1), n^4\}) \\
& =  \E \big[  {S}^2_K(x_1) - ( {S}^2_K(x_1)-n^4)I( {S}^2_K(x_1)>n^4)  \big]\\
& \le \E {S}^2_K(x_1) \\
& \lesssim  \log n,
\end{align*}
and 
\begin{align*}
{\E}x_1^2+({\E}|x_1|)^2 -2 \E{\delta}(x_1)\E |x_1|  &\le \V(x_1)+(\E |x_1|)^2 +(\sqrt{2\log n})^2 \lesssim \log n
\end{align*}
Then we have
\begin{align*}
\tilde{\sigma}^2_x &\le\V({S}_K(x_1)){P}(\sqrt{2\log n} <|x_2|\le 2\sqrt{2 \log n}) + \V(|x_1|) {P}(|x_2|>2\sqrt{2 \log n}) \\
&\quad +(\E {\delta}(x_1))^2 +(\E |x_1|)^2{P}(|x_2|>2\sqrt{2 \log n})-2\E {\delta}(x_1)\E |x_1|{P}(|x_2|>2\sqrt{2 \log n})\\
&\le {\E}{S}^2_K(x_1)n^{-1}+(\E {\delta}(x_1))^2+[{\E}x_1^2+({\E}|x_1|)^2 -2\E {\delta}(x_1)\E |x_1|] \cdot \frac{1}{n \sqrt{\pi\log n}}\\
& \lesssim n^{-1/2}2^{8K}M_n^2 K^2 \\
& \lesssim \frac{\log n}{\sqrt{n}}
\end{align*}

\subsection{Proof of Proposition 3}

We only prove the proposition for $\theta$. The argument for $\mu$ is the same. We need the following lemma for the proof.

\bel \label{moments.s.lem2}
Let $X\sim N(\theta,1)$ and ${S}_K(x) = \sum_{k=1}^K g_{2k}M_n^{-2k+1}H_{2k}(x)$ with $M_n = 8\sqrt{\log n}$ and $K = r\log n$ for some $r>0$. Then for all $|\theta| \le 4\sqrt{2 \log n}$,
\[
\bigg|\E {S}_K(X) -{|\theta|} \bigg| \le \frac{4M_n}{\pi(2K+1)}\lesssim \frac{1}{\sqrt{\log n}},
\]
and  $\E {S}^2_K(X) \lesssim n^{6r}\log^3 n$.
\eel

\paragraph{Case 1.} $\theta = 0$. Note that ${\delta}(x_1)$ can be written as
\[
{\delta}(x_1) = {S}_K(x_1) - ({S}_K(x_1)-n^2)I({S}_K(x_1)\ge n^2).
\]
Consequently,
\begin{align*}
| B_x | & \le | \E {S}_K(x_1) | +\E \{({S}_K(x_1)-n^2)I({S}_K(x_1)\ge n^2)\} \\
&\quad +(|\E {S}_K(x_1)|+ \E |x_1| ) \mathbb{P}(|x_2|>2\sqrt{2 \log n}) \\
& \equiv B_1+B_2+B_3.
\end{align*}
By definition of  ${S}_K(x_1)$ we have
\begin{equation}
B_1 = 0.
\end{equation}
It follows from the standard bound for normal tail probability $\Phi(-z) \le z^{-1} \phi(z)$ for $z>0$ that
\begin{equation}
\mathbb{P}(|x_2| >2\sqrt{2\log n}) = 2\Phi(-2\sqrt{2\log n}) \le \frac{1}{2\sqrt{\pi\log n}} n^{-4}.
\end{equation}
And in this case 
\begin{equation}
\E |x_1| = 2\phi(0).
\end{equation}
It then follows that
\begin{equation}
B_3 \le  2\phi(0) \cdot \frac{1}{2\sqrt{\pi\log n}} n^{-4} = \frac{1}{\pi \sqrt{2\log n}}n^{-4}.
\end{equation}
Now consider $B_2$. Note that for any random variable $X$ and any constant $\lambda >0$,
\[
\E (XI(X\ge \lambda))\le \lambda^{-1} \E(X^2 I(X\ge\lambda) \le \lambda^{-1} \E X^2.
\]
This together with Lemma 4 yields that
\begin{equation} \label{b2}
B_2\le \E \{({S}_K(x_1)I({S}_K(x_1)\ge n^2)\} \le  n^{-2} \E({S}^2_K(x_1)) \lesssim  n^{6r-2}\log^3 n.
\end{equation}
Combining the three pieces together, we have
\[
|B_x| \le B_1+B_2+B_3 \lesssim n^{6r-2} \log^3 n.
\]
We now consider the variance. It follows from Lemma \ref{var.lem2} and Lemma \ref{moments.s.lem2} that
\begin{align*}
\tilde{\sigma}^2_x &\le\V({S}_K(x_1)){P}(|x_2|\le 2\sqrt{2 \log n}) + \V(|x_1|) {P}(|x_2|>2\sqrt{2 \log n}) \\
&\quad +[(\E {\delta}(x_1))^2 +(\E |x_1|)^2]{P}(|x_2|>2\sqrt{2 \log n})-2\E {\delta}(x_1)\E |x_1|{P}(|x_2|>2\sqrt{2 \log n})\\
&\le {\E}{S}^2_K(x_1)+[(\E {\delta}(x_1))^2+({\E}|x_1|)^2 -2\E {\delta}(x_1)\E |x_1|] \cdot \frac{1}{2 n^4 \sqrt{\pi\log n}}\\
& \lesssim n^{6r}\log^3 n
\end{align*}
as long as $r<3/4$.

\paragraph{Case 2.} $0< | \theta| \le \sqrt{2\log n}$. In this case
\begin{align*}
|B_x| &= | \E({\delta}(x_1)) {P}(|x_2|\le 2\sqrt{2 \log n}) + \E(|x_1|) {P}(|x_2|>2\sqrt{2 \log n}) -| \theta || \\
&\le | \E ({\delta}(x_1)) -| \theta| |+ |\E |x_1| -|\theta| | {P}(|x_2|>2\sqrt{2 \log n}) \\
&\lesssim | \E{S}_K(x_1)-| \theta| | + \E \{({S}_K(x_1)-n^2)I({S}_K(x_1)\ge n^2)\}  + n^{-4}\\
& \lesssim 1/\sqrt{\log n} + n^{6r-2}\log^3 n + n^{-4}\\
&\lesssim 1/\sqrt{\log n}
\end{align*}
as long as $r<1/3$.
Similarly, note that
\[
\E {\delta}(x_1) = \E {S}_K(x_1) + B_2
\]
then
\begin{align*}
(\E {\delta}(x_1)-\E |x_1|)^2  &\le 2(\E {\delta}(x_1))^2 +2(\E |x_1|)^2 \\
&\le 2( \E {S}_K(x_1) +B_2)^2 + 2(|\theta|+1)^2 \\
&\lesssim n^{6r}\log^3 n.
\end{align*}
Hence the variance can be bounded as follows.
\begin{align*}
\tilde{\sigma}_x^2 &\le  \E {S}^2_K(x_1) + \E x_1^2  \mathbb{P}(|x_2|>2\sqrt{2 \log n})\\
&\quad  +(\E {\delta}(x_1)-\E |x_1|)^2 \mathbb{P}(|x_2|>2\sqrt{2 \log n})\\
& \lesssim n^{6r}\log^3n.
\end{align*}

\paragraph{Case 3.} $\sqrt{2\log n}< | \theta| \le 4\sqrt{2\log n}$. In this case
\begin{align*}
|B_x| &= | \E({\delta}(x_1)) {P}(|x_2|\le 2\sqrt{2 \log n}) + \E(|x_1|) {P}(|x_2|>2\sqrt{2 \log n}) -| \theta || \\
&\le | \E ({\delta}(x_1)) -| \theta| |+ |\E |x_1| -|\theta| | \\
&\lesssim | \E{S}_K(x_1)-| \theta| | + \E \{({S}_K(x_1)-n^2)I({S}_K(x_1)\ge n^2)\}  + 2\phi(\theta)\\
& \lesssim 1/\sqrt{\log n} + n^{6r-2}\log^3 n + n^{-1}\\
&\lesssim 1/\sqrt{\log n}
\end{align*}
as long as $r<1/3$. The variance can be bounded similar to the Case 2.

\paragraph{Case 4.} $|\theta| >4\sqrt{2\log n}$. In this case, same argument follows from the proof of Case 4 in Proposition 1.
\qed

\paragraph{Proof of Proposition 4.}
Similar to the proofs of previous propositions, we only prove the statements about $\theta$. The argument for $\mu$ is the same.
\paragraph{Case 1.} $\theta = 0$. 
Note that $| B_x | = \E(|x_1|) {P}(|x_2|>2\sqrt{2 \log n}).$
It follows from the standard bound for normal tail probability $\Phi(-z) \le z^{-1} \phi(z)$ for $z>0$ that
\begin{equation}
{P}(|x_2| >2\sqrt{2\log n}) = 2\Phi(-2\sqrt{2\log n}) \le \frac{1}{2\sqrt{\pi\log n}} n^{-4}.
\end{equation}
And in this case $E |x_1| = 2\phi(0)$.
It then follows that
\begin{equation}
|B_x| \le  2\phi(0) \cdot \frac{1}{2\sqrt{\pi\log n}} n^{-4} = \frac{1}{\pi \sqrt{2\log n}}n^{-4}.
\end{equation}
We now consider the variance. It follows that
\begin{align*}
\tilde{\sigma}^2_x &\le \V(|x_1|) {P}(|x_2|>2\sqrt{2 \log n}) \le {\E}x_1^2\cdot \frac{1}{2 n^4 \sqrt{\pi\log n}} \lesssim \frac{1}{n^4\sqrt{\log n}}.
\end{align*}

\paragraph{Case 2.} $0< | \theta| \le \sqrt{2\log n}$. In this case
\[
|B_x| = \big|\E|x_1|P(|x_2|>2\sqrt{2\log n})-\theta \big|\le \E|x_1|P(|x_2|>2\sqrt{2\log n})+|\theta|.
\]
The standard bound for normal tail probability yields
\[
{P}(|x_2| > 2\sqrt{2\log n}) \le 2\Phi(-\sqrt{2\log n}) \le \frac{1}{\sqrt{\pi \log n}}n^{-1}
\]
Note that
\[
\E |x_1| = |\theta| +2\phi(\theta) - 2|\theta|\Phi(-|\theta|)\le|\theta| +1\le\sqrt{2\log n}+1.
\]
Then we have
\[
|B_x| \le \bigg(\sqrt{2\log n}+1  \bigg)\cdot \frac{1}{\sqrt{\pi \log n}}n^{-1} +|\theta|\le 3 n^{-1}+\sqrt{2\log n}.
\]
On the other hand, since
\begin{align*}
{\E}x_1^2 &\le \V(x_1)+(\E |x_1|)^2  \le 1 + 2 \log n.
\end{align*}
we have
\begin{align*}
\tilde{\sigma}^2_x &\le \V(|x_1|){P}(|x_2|>2\sqrt{2 \log n}) \le {\E}x_1^2\cdot \frac{1}{ n \sqrt{\pi\log n}} \lesssim \frac{\sqrt{\log n}}{n}.
\end{align*}

\paragraph{Case 3} $\sqrt{2\log n} \le |\theta| \le 4\sqrt{2\log n}$. In this case \ref{moments.s.lem},
\begin{align*}
|B_x| &= | \E(|x_1|) {P}(|x_2|>2\sqrt{2 \log n}) -| \theta || \\
&\le  |\E |x_1| -|\theta| | +|\theta| {P}(|x_2|\le 2\sqrt{2\log n})\\
&\lesssim \sqrt{\log n}.
\end{align*}
For the variance, similarly we have
\begin{align*}
\tilde{\sigma}^2_x\le {\E}x_1^2 \lesssim \log n.
\end{align*}

\paragraph{Case 4.} $|\theta| >4\sqrt{2\log n}$. In this case, the standard bound for normal tail probability yields that
\[
{P}(|x_2| \le 2\sqrt{2\log n}) \le 2\Phi(-(|\theta|-2\sqrt{2\log n})) \le 2\Phi\bigg( -\frac{|\theta|}{2} \bigg) \le \frac{4}{|\theta|}\phi\bigg( \frac{|\theta|}{2} \bigg).
\]
In particular, 
\[
{P}(|x_2| \le 2\sqrt{2\log n})  \le 2\Phi(-2\sqrt{2\log n}) \le \frac{1}{2\sqrt{\pi \log n}}n^{-4}.
\]
Hence,
\begin{align*}
&|B_x| \le | \E(|x_1|) {P}(|x_2|>2\sqrt{2 \log n}) -| \theta || \le |\E |x_1| -|\theta| | +\E|x_1|  {P}(|x_2| \le 2\sqrt{2\log n})\\
& \le 2 \phi(\theta) +(|\theta|+1) {P}(|x_2|\le2\sqrt{2 \log n})  \le 2 \phi(\theta) + 4\phi\bigg(\frac{|\theta|}{2}\bigg)  \lesssim \frac{1}{n^2}.
\end{align*}
For the variance, similarly we have
\begin{align*}
\tilde{\sigma}^2_x &\le \V(|x_1|)+(\E |x_1|)^2{P}(|x_2|\le2\sqrt{2 \log n})  \\
&= 1+o(1).
\end{align*}
\qed

\paragraph{Proof of Proposition 5.}
In this case
\[
|B_x| = \big|\E|x_1|P(|x_2|>2\sqrt{2\log n})-\theta \big|\le \E|x_1|P(|x_2|>2\sqrt{2\log n})+|\theta|.
\]
The standard bound for normal tail probability yields
\[
{P}(|x_2| > 2\sqrt{2\log n}) \le 2\Phi(-\sqrt{2\log n}) \le \frac{1}{\sqrt{\pi \log n}}n^{-1}
\]
Note that
\[
\E |x_1| = |\theta| +2\phi(\theta) - 2|\theta|\Phi(-|\theta|)\le|\theta| +1\le L_n+1.
\]
Then we have
\[
|B_x| \le \bigg(L_n+1  \bigg)\cdot \frac{1}{\sqrt{\pi \log n}}n^{-1} +|\theta|\lesssim L_n.
\]
On the other hand, since
\begin{align*}
{\E}x_1^2 &\le \V(x_1)+(\E |x_1|)^2  \le 1+(1 + L_n)^2.
\end{align*}
we have
\begin{align*}
\tilde{\sigma}^2_x &\le \V(|x_1|){P}(|x_2|>2\sqrt{2 \log n}) \le {\E}x_1^2\cdot \frac{1}{ n \sqrt{\pi\log n}} \lesssim \frac{\sqrt{\log n}}{n}.
\end{align*}
\section{Proofs of Technical Lemmas}

\paragraph{Proof of Lemma \ref{constant.lem}.} By Lemma 2 of \cite{cai2011testing}, for $x\in [-1,1]$, we have
\[
\max_{x \in [-1,1]} |G_{K}(x) - |x| |\le \frac{2}{\pi (2K+1)}.
\]
Then for $x' \in [-M_n, M_n]$, we have $x' = M_nx$, and thus
\begin{equation} \label{diff1}
\max_{x'} |\tilde{G}_{K}(x') - |x'| |\le \frac{2M_n}{\pi (2K+1)}.
\end{equation}
Set $x'=0$, we obtain the statement.
\qed

\paragraph{Proof of Lemma \ref{moments.s.lem}.}
The first statement follows from Lemma 2 in \cite{cai2011testing} and that
\begin{align}
\big|\E {S}_K(X) -|\theta| \big| &\le \bigg|\sum_{k=0}^Kg_{2k}M_n^{-2k+1}\theta^{2k} -{|\theta|} \bigg| +  \frac{2M_n}{\pi (2K+1)} \\
& \le \frac{4M_n}{\pi (2K+1)}.
\end{align}
To bound $\E{S}^2_K(X)$, it follows from Lemma 3 in \cite{cai2011testing} and that
\begin{align}
\E {S}^2_K(X) &\le \bigg( \sum_{k=1}^K|g_{2k}| M_n^{-2k+1}(\E H_{2k}^2(X))^{1/2} \bigg)^2 \\
& \le 2^{6K} \bigg( \sum_{k=1}^KM_n^{-2k+1}(2M_n^2)^k \bigg)^2 \\
& \le 2^{8K}M_n^2 K^2 
\end{align}
\qed

\paragraph{Proof of Lemma \ref{var.lem2}.} For equation (\ref{var21}), since events $A$ and $B$ are independent of random variables $X$ and $Y$, we have
\begin{align*}
\V(XI(A) + YI(B)) &= \E\big[X^2 I(A) +Y^2I(B) +2XYI(A)I(B)   \big] -\big( \E XP(A)+\E Y P(B) \big)^2\\
&= \E X^2P(A) +\E Y^2P(B) -(\E X)^2 P^2(A)-(\E Y)^2P^2(B)\\
&\quad -2\E X\E YP(A)P(B) \\
&= \V(X)P(A)+\V(Y) P(B) +(\E X)^2(P(A)-P^2(A))\\
&\quad+(\E Y)^2(P(B)-P^2(B))-2\E X\E YP(A)P(B)
\end{align*}
Equation (\ref{var22}) follows directly from the above derivation.
\qed

\paragraph{Proof of Lemma \ref{moments.s.lem2}.}
It follows from Lemma 2 in \cite{cai2011testing} that
\begin{align}
\bigg|\E {S}_K(X) - {|\theta|} \bigg| &=\bigg|\sum_{k=0}^Kg_{2k}M_n^{-2k+1}{\theta}^{2k} -{|\theta|}  \bigg| +  \frac{2M_n}{\pi (2K+1)} \\
& \le \frac{4M_n}{\pi (2K+1)}.
\end{align}
To bound $\E {S}^2_K(X)$, it follows that when $K=r\log n$ for some $r>0$,
\begin{align}
\E {S}^2_K(X) &\le \bigg( \sum_{k=1}^K|g_{2k}| M_n^{-2k+1}(E H_{2k}^2(X))^{1/2} \bigg)^2 \\
& \le 2^{6K} \bigg( \sum_{k=1}^KM_n^{-2k+1}(2M_n^2)^k \bigg)^2 \\
& \le 2^{8K}M_n^2 K^2 \\
&\lesssim n^{6r} \log^3 n.
\end{align}
\qed

\appendix
\counterwithin{figure}{section}
\counterwithin{table}{section}

\section{Supplementary Theoretical Discussions}

\subsection{Minimax Optimal Rate When $\beta=1/2$}

When $\beta=1/2$, we consider the following estimator
\[
\widetilde{T}_0={2}\sum_{i=1}^n \hat{U}_i(x_i)\hat{U}_i(y_i),
\]
where 
\[
\hat{U}_i(x_i)= (|x_{i1}|-\alpha )\cdot I(|x_{i2}|> \sqrt{2\log 2}),
\]
and
\[
\alpha = \frac{\E [|\xi| I(\xi^2>2\log 2)]}{P(\xi^2>2\log 2)},\qquad \text{for $\xi\sim N(0,1)$.}
\]
Following the similar arguments as in the proof of Theorem 4 of \cite{collier2020estimation} as well as the proof of Theorem 2 of the main paper, it can be shown that
\beq
\sup_{\substack{(\theta,\mu,\Sig_1,\Sig_2) \in D^\infty(s,L_n)}}\mathcal{R}(\widetilde{T}_0 ) \lesssim \frac{s^2(L_n^2+1)}{n^2}.
\eeq
Comparing the above risk upper bound to the minimax lower bound in Theorem 1, we have, for  $\beta=1/2$ and $L_n\ge\sqrt{\log(1+n/s^2)}\asymp 1$,
\beq
\inf_{\hat{T}} \sup_{\substack{(\theta,\mu,\Sig_1,\Sig_2) \in D_0^\infty(s,L_n)}}\mathcal{R}(\hat{T}) \asymp \frac{s^2L_n^2}{n^2}.
\eeq

\subsection{Complexities from the Covariances}

In the main paper, most of our theoretical results are obtained under the assumption $\Sig_1=\Sig_2={\bf I}$, whereas our Theorem 4 essentially take the worst case over these covariances. In this section, we discuss the cases with known and unknown covariances and the corresponding estimators.

\paragraph{Known covariances.} On the one hand, if the diagonals of the covariance matrices are all 1's, while the off-diagonal entries are known and possibly nonzero, then in principle our proposed estimators can still be applied, although the rate of convergence might not remain the same, nor does the minimax optimality property. Nevertheless, the analysis of these estimators can be technically challenging. For example, obtaining the risk upper bound of $\widehat{T}_K$ (or $\widehat{T}^S_K$) requires calculation of the covariances between the hybrid components $\hat{V}_{i,K}(x_i)$ (or $\hat{V}^S_{i,K}(x_i)$) for correlated $x_i$'s, which relies on properties of Hermite polynomials. The final risk upper bounds will depend on the specific covariance structure. 

On the other hand, if the diagonals are not 1, then an extra rescaling step is needed in our construction of the polynomial approximation based estimators. Specifically, suppose $\Sig_1$ and $\Sig_2$ have diagonal entries $\{\sigma_i^2\}_{i=1}^n$ and $\{\tau_i^2\}_{i=1}^n$, respectively, then we can define the adjusted version of $\widehat{T}_K$ as 
\[
\widehat{T}'_K = {2}\sum_{i=1}^n \hat{V}'_{i,K}(x_i)\hat{V}'_{i,K}(y_i),
\]
where
\[
\hat{V}'_{i,K}(x_i)= {\delta}'_K(x_{i1})I(|x_{i2}|\le 2\sigma_i\sqrt{2\log n}) + |x_{i1}| I(|x_{i2}|> 2\sigma_i\sqrt{2\log n}),
\]
\[
\hat{V}'_{i,K}(y_i)= {\delta}'_K(y_{i1})I(|y_{i2}|\le 2\tau_i\sqrt{2\log n}) + |y_{i1}| I(|y_{i2}|> 2\tau_i\sqrt{2\log n}),
\]
and
\[
{\delta}'_K(x_i) =\sigma_i \min\{{S}_K(x_i/\sigma_i) ,n^2 \},\qquad {\delta}'_K(y_i) =\tau_i \min\{{S}_K(y_i/\tau_i) ,n^2 \}.
\]
Similarly, one can define the adjusted version of $\widehat{T}_K^S$ that takes into account the knowledge of the variances.

\paragraph{Unknown variances.} When the covariances are completely unknown, the estimation problem will be extreme difficult since one may not be able to distinguish the mean from the variance components based on the observed data. Therefore, in the following, we only consider the cases where the diagonals of $\Sig_1$ and $\Sig_2$ are unknown but identical, say, $\Sig_1=\sigma_1^2{\bf I}$ and $\Sig_2=\sigma_2^2{\bf I}$, respectively.

In this case, based on the previous discussions, especially the definitions of the adjusted estimators $\widehat{T}'_K$,  it is important to estimate the variances $\sigma_1^2$ and $\sigma_2^2$. Toward this end, if in addition we assume that both $\theta$ and $\mu$ are sparse vectors in themselves, we may take advantage of such sparsity and estimate $\sigma^2_k$ by the smaller observations since they are likely to correspond to mean zero Gaussian random variables. Following the ideas in \cite{collier2017minimax,collier2018optimal}, we may estimate $\sigma_1$ and $\sigma_2$ by
\beq \label{sigma.estr1}
\hat{\sigma}_{1} =9 \sqrt{\frac{1}{\lfloor n/2 \rfloor}\sum_{j\le n/2}x_{(j)}^2 },\quad \quad \hat{\sigma}_{2} =9 \sqrt{\frac{1}{\lfloor n/2 \rfloor}\sum_{j\le n/2}y_{(j)}^2 },
\eeq
where $x_{(1)} \le x_{(2)} \le ... \le x_{(n)}$ and $y_{(1)} \le y_{(2)} \le ... \le y_{(n)}$ are ordered statistics associated to $\bold{x}_n$ and $\bold{y}_n$. With the above variance estimators, we can estimate the T-score using the above adjusted estimators by plugging-in the variance estimators.

Moreover, when $\beta\in(0,1/2)$, there is an extra advantage of the simple thresholding estimator $\tilde{T}$. Specifically, if we define $D_{\sigma}^\infty(s,L_n) = \big\{(\theta,\mu,\Sig_1,\Sig_2):  (\theta,\mu)\in D(s),\max(\|\theta\|_\infty , \|\mu\|_\infty )\le L_n,  \Sig_1=\Sig_2=\sigma^2{\bf I}_n \big\}$, then by the proofs of Theorems 1 and 3, it can be shown that
\beq
\inf_{\hat{T}}\sup_{\substack{(\theta,\mu,\Sig_1,\Sig_2) \in D_{\sigma}^\infty(s,L_n)}}\mathcal{R}(\hat{T} ) \gtrsim\frac{s^2L_n^2}{n^2}\cdot\min\{\sigma^2 \log n,L_n^2\},
\eeq
and
\beq
\sup_{\substack{(\theta,\mu,\Sig_1,\Sig_2) \in D_{\sigma}^\infty(s,L_n)}}\mathcal{R}(\widetilde{T} ) \lesssim \frac{\sigma^2(L_n^2+\sigma^2\log n)s^2\log n}{n^2}.
\eeq
Again, if in addition $L_n\le \sigma\sqrt{2\log n}$, then
\beq
\sup_{\substack{(\theta,\mu,\Sig_1,\Sig_2) \in D_\sigma^\infty(s,L_n)}}\mathcal{R}(\widetilde{T} ) \lesssim \frac{s^2L_n^4}{n^2}+\frac{\sigma^4\log^2 n}{n^3}+\frac{\sigma^2L_n^2\log n}{n^2}.
\eeq
Therefore, whenever $L_n\gtrsim \sigma$, $\widetilde{T}$ is minimax optimal with the optimal rate of convergence being $\frac{L_n^2s^2}{n^2}\min\{\sigma^2\log n,L_n^2\}.$

\section{Supplementary Figures and Tables from Numerical Studies}

\paragraph{Supplementary simulation results.} In our Section 3, to generate dependent observations from a non-identity covariance matrix, we considered the block-wise covariances where $\Sig$ is block diagonal where each block is either a $10\times 10$ Toeplitz matrix or a $1000\times 1000$ exchangeable covariance matrix whose off-diagonal elements are $0.5$. In particular, the $10\times 10$ Toeplitz matrix is given as follows
\[
\begin{bmatrix}
1&0.9&0.8&0.7&0.6&0.5&...&0.1\\
0.9&1&0.9&0.8&0.7&0.6&...&0.2\\
0.8&0.9&1&0.9&0.8&0.7&...&0.3\\
...\\
0.1&0.2&0.3&0.4&0.5&0.6&...&1
\end{bmatrix}.
\]
Tables \ref{table:t2} and \ref{table:t3} include the empirical \textsc{rmse} of the five estimators under covariance structures $\Sig_1$ and $\Sig_2$, respectively.

\begin{table}[h!]
	\caption{Empirical \textsc{rmse} under covariance $\Sig_1$.}
	\vskip .2cm
	\centerline{\tabcolsep=3truept\begin{tabular*}{ 0.78 \textwidth}{cc|ccccc|ccccc}
			\hline
			$\frac{n}{10^4}$&   $s$  &  $\widehat{T}^{S*}_K$  & $\widehat{T}^S_K$   & 	$\widetilde{T}$   &      $\widehat{T}_K$  & $\overline{T} $&$\widehat{T}^{S*}_K$  & $\widehat{T}^S_K$   & 	$\widetilde{T}$   &      $\widehat{T}_K$  & $\overline{T}$ \\  
			\hline  
			&&	\multicolumn{5}{c}{Sparse Pattern I} &\multicolumn{5}{c}{Sparse Pattern II}\\
			&50& 11.58 & 22.53 & 25.99 & 23.25 & 1909.3&7.89 & 26.97 & 36.25 & 31.72 & 1908.5  \\
			&100& 10.16 & 21.44 & 28.93 & 26.60 & 954.7&7.82 & 26.55 & 36.19 & 31.98 & 953.9 \\
			15   & 200&  10.74 & 23.10 & 31.72 & 28.66 & 476.0&7.23 & 25.19 & 34.72 & 30.5 & 477.4\\
			& 400  & 10.11 & 22.14 & 30.75 & 27.67 & 237.9&9.24 & 25.16 & 30.28 & 26.92 & 237.7\\
			& 800&10.24 & 21.56 & 28.53 & 25.76 & 118.8&8.44 & 26.04 & 34.06 & 30.19 & 118.6 \\
			\hline  
			& 50& 11.47 & 24.65 & 30.30 & 27.37 & 3821.8& 10.53 & 28.01 & 37.42 & 32.88 & 3821.1 \\
			&100& 10.82 & 22.60 & 31.32 & 27.92 & 1910.8&8.87 & 26.51 & 35.32 & 30.96 & 1909.7 \\
			30    & 200& 11.01 & 23.15 & 28.85 & 26.01 & 954.0& 8.63 & 28.06 & 35.27 & 30.83 & 953.5\\
			& 400  & 10.99 & 22.63 & 29.89 & 26.55 & 476.9&10.52 & 26.39 & 31.89 & 28.19 & 476.9\\
			& 800&  10.79 & 22.71 & 30.20 & 26.85 & 237.9& 9.57 & 26.66 & 33.2 & 29.25 & 238.0\\
			\hline  
			& 50& 10.94 & 26.16 & 35.45 & 31.23 & 6364.9&9.19 & 30.27 & 39.04 & 34.17 & 6366.3\\
			&100&  11.62 & 23.04 & 28.53 & 25.68 & 3182.8&11.36 & 26.92 & 31.66 & 27.97 & 3184.7 \\
			50    & 200&  11.25 & 22.29 & 29.01 & 25.83 & 1590.6&10.84 & 27.39 & 31.60 & 27.90 & 1591.0\\
			& 400  &10.61 & 23.38 & 30.48 & 26.77 & 795.0&9.66 & 27.93 & 34.07 & 29.93 & 795.4\\
			& 800& 11.38 & 23.25 & 29.63 & 26.35 & 397.4&10.25 & 27.14 & 32.64 & 28.76 & 397.2\\
			\hline
	\end{tabular*}}
	\label{table:t2}
\end{table}

\begin{table}[h!]
	\centering
	\caption{Empirical \textsc{rmse} under covariance $\Sig_2$.}
	\vskip .2cm
	\centerline{\tabcolsep=3truept\begin{tabular*}{ .78 \textwidth}{cc|ccccc|ccccc}
			\hline
			$\frac{n}{10^4}$&   $s$  &  $\widehat{T}^{S*}_K$  & $\widehat{T}^S_K$   & 	$\widetilde{T}$   &      $\widehat{T}_K$  & $\overline{T} $&$\widehat{T}^{S*}_K$  & $\widehat{T}^S_K$   & 	$\widetilde{T}$   &      $\widehat{T}_K$  & $\overline{T}$ \\  
			\hline  
			&&	\multicolumn{5}{c}{Sparse Pattern I} &\multicolumn{5}{c}{Sparse Pattern II}\\ 
			&50& 10.59 & 23.56 & 31.26 & 28.15 & 1913.9&7.27 & 23.89 & 38.89 & 34.82 & 1913.7\\
			&100&  9.69 & 24.12 & 33.90& 30.13 & 950.3&9.35 & 26.58 & 34.95 & 31.06 & 949.9 \\
			15   & 200& 10.06 & 23.47 & 30.95 & 27.58 & 477.5&8.15 & 24.46 & 33.50 & 29.69 & 482.1 \\
			& 400  & 10.12 & 21.66 & 29.35 & 26.38 & 239.5&7.84 & 25.80 & 34.89 & 30.93 & 238.2 \\
			& 800& 9.64 & 21.88 & 30.18 & 27.14 & 118.5& 8.49 & 25.51 & 33.03 & 29.18 & 118.8\\
			\hline  
			& 50&  9.23 & 25.31 & 35.58 & 31.36 & 3818.3 & 8.66 & 25.21 & 33.44 & 29.05 & 3826.3\\
			&100& 12.29 & 23.97 & 31.05 & 27.91 & 1908.5& 9.57 & 26.84 & 35.01 & 30.96 & 1914.1\\
			30   & 200&  11.38 & 22.50 & 29.33 & 26.13 & 958.0& 9.97 & 26.44 & 31.81 & 28.05 & 953.7\\
			& 400  &11.63 & 23.15 & 29.64 & 26.57 & 477.1& 8.88 & 26.18 & 34.72 & 30.77 & 476.8\\
			& 800&  10.71 & 22.93 & 30.40 & 27.10 & 237.6&9.37 & 26.48 & 33.02 & 29.07 & 237.6\\
			\hline  
			& 50&8.97 & 25.34 & 33.31 & 29.20 & 6345.1&9.35 & 28.84 & 35.80 & 31.22 & 6360.9 \\
			& 100& 12.43 & 23.16 & 26.86 & 24.02 & 3178.4&8.55 & 28.53 & 37.41 & 32.49 & 3178.0  \\
			50  & 200&  10.97 & 22.22 & 28.75 & 25.66 & 1586.2&8.90 & 28.15 & 35.82 & 31.19 & 1589.8 \\
			& 400  &11.07 & 22.40 & 28.48 & 25.34 & 793.8&9.91 & 26.84 & 32.92 & 28.94 & 797.0 \\
			& 800& 10.80 & 23.25 & 30.33 & 26.93 & 397.7&10.26 & 27.14 & 33.88 & 29.96 & 398.2 \\
			\hline
	\end{tabular*}}
	\label{table:t3}
\end{table}

\begin{figure}[htp]
	\hspace{-3cm}
	\includegraphics[angle=0,width=21cm]{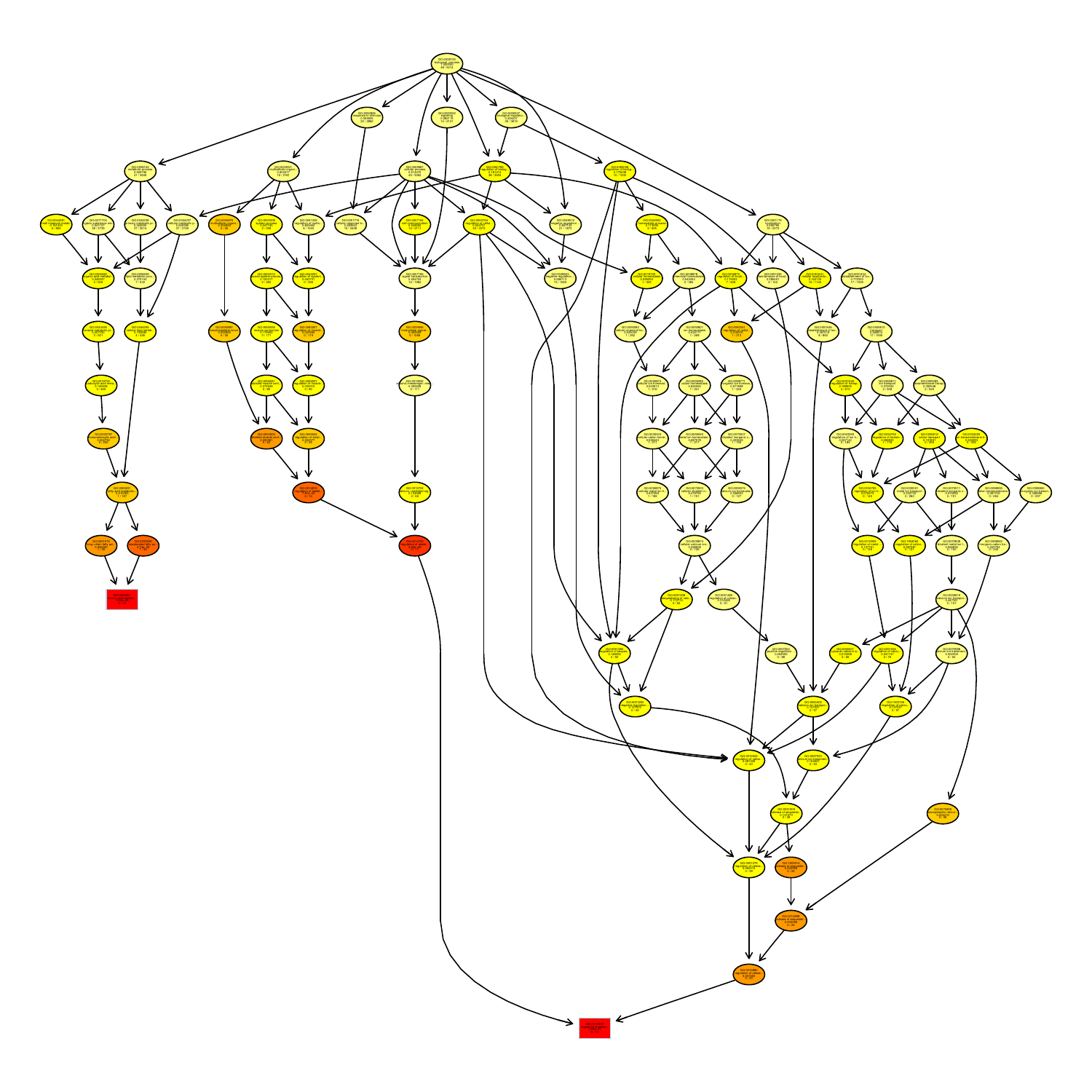}
	\caption{Directed acyclic graph of \textsc{go} biological processes connected by some path to the most significant processes  from the \textsc{gsea} analysis. Yellow: least significant; Red: most significant; Rectangles: top \textsc{gsea} results.}\label{GO.BP}
\end{figure}

\paragraph{Real data analysis.} In our gene set enrichment analysis, 5,023 biological processes from Gene Ontology (\textsc{go}) (\citealt{botstein2000gene}) that contain at least 10 genes were tested.
Figure  \ref{GO.BP}  presents the directed acyclic graphs of the \textsc{go} biological processes  that linked to  the most significant \textsc{go} terms  from the simultaneous signal \textsc{gsea} analysis. 

\label{lastpage}

\end{document}